\documentclass{jfm}
\usepackage{graphicx}
\usepackage{float}
\usepackage{rotating}
\usepackage[flushleft]{threeparttable}
\usepackage{epstopdf,epsfig}
\usepackage{lscape}
\usepackage{amsmath}
\usepackage{mathtools}
\usepackage{tabularx}
\usepackage{tikz}
\usepackage{centernot}
\usepackage{booktabs}
\usepackage{comment} 
\usepackage{subcaption}
\usepackage{siunitx}
\usepackage{lineno}
\usepackage{units}
\usepackage{nicefrac,xfrac}
\usepackage{soul}
\usepackage{ulem} % to strike things out with \sout
\usetikzlibrary{shapes.geometric}
\usepackage{flafter}
\usepackage{hyphenat}
\usepackage{multicol}
\usepackage{multirow}
\usepackage{xr}
\usepackage{bm}
\usepackage{enumitem}
\newlist{myenumi}{description}{10}
\setlist[myenumi]{leftmargin=23pt,itemsep=2pt,topsep=2pt,parsep=2pt}
\newlist{myenumi2}{description}{10}
\setlist[myenumi2]{leftmargin=40pt,itemsep=2pt,topsep=2pt,parsep=2pt}

\usepackage{ragged2e}
\newcolumntype{P}[1]{>{\RaggedRight\hspace{0pt}}p{#1}}

\newcommand{\LL}{\textsf{L}}
\newcommand{\HH}{\textsf{H}}

\newcommand{\II}{\textsf{I}}
\newcommand{\TT}{\textsf{T}}

\renewcommand{\d}{\, \mbox{d}}
\providecommand\bnabla{\boldsymbol{\nabla}}
\newcommand{\uu}{\bm{u}}
\newcommand{\xx}{\bm{x}}

\newcommand{\Rigbb}{\overline{\overline{Ri_g}}}

\shorttitle{Experimental properties of stratified turbulence}
\shortauthor{A. Lefauve \& P. F. Linden}

\title{\Large{Experimental properties of continuously-forced, shear-driven,  stratified turbulence.} \\ Part 1. Mean flows, self-organisation, turbulent fractions.}

\author{Adrien Lefauve and P. F. Linden}
\affiliation{Department of Applied Mathematics and Theoretical Physics, University of Cambridge \\ Centre for Mathematical Sciences, Wilberforce Road, Cambridge, CB3 0WA, UK.}

\begin{document}

\maketitle

%\tableofcontents

%\clearpage

% Current estimated length of papers (split into Part 1 and Part 2):

% \textbf{Part 1.  Mean flows, turbulent fraction, and self-organisation}: $1+1.5+2.5+4.5+5.5+3+7.5+1=26.5$ pages $+ 2$ (appendix A and B) $+2$ (refs) $=30$ (total)

\begin{abstract}

We study the experimental properties of  exchange flows in a stratified inclined duct (SID), which are simultaneously turbulent, strongly stratified by a mean vertical density gradient, driven by a mean vertical shear, and continuously forced by gravity. We focus on the `core' shear layer away from the duct walls, where these flows are excellent experimentally-realisable approximations of canonical hyperbolic-tangent stratified shear layers, whose forcing allows mean and turbulent properties to reach quasi steady states.  We analyse  state-of-the-art data sets of the time-resolved density and velocity in three-dimensional sub-volumes of the duct in 16 experiments covering a range of flow regimes (Holmboe waves, intermittent turbulence, full turbulence). In this Part 1 we first reveal the permissible regions in the multi-dimensional parameter space (Reynolds number, bulk Richardson number, velocity-to-density layer thickness ratio), and their link to experimentally-controllable parameters.  Reynolds-averaged balances then reveal the subtle momentum forcing and dissipation mechanisms in each layer, the broadening or sharpening of the density interface, and the importance of the streamwise non-periodicity of these flows. Mean flows suggest a tendency towards self-similarity of the velocity and density profiles with increasing turbulence, and gradient Richardson number statistics support prior `internal mixing' theories of  `equilibrium Richardson number', `marginal stability' and `self-organised criticality'. Turbulent volume fractions based on enstrophy and overturn thresholds quantify the nature of turbulence between different regimes in different regions of parameter space, while highlighting the challenges of obtaining representative statistics in spatio-temporally intermittent flows. These insights may stimulate and assist the development of numerical simulations with a higher degree of experimental  realism.
\end{abstract}

% \begin{keywords}
% \end{keywords}

%%%%%%%%%%%%%% INTRODUCTION %%%%%%%%%%%%%%%%%%%%

\section{Introduction}\label{sec:intro}

In this two-part study we present experimental results relevant to a wide class of geophysical flows that are simultaneously:

\begin{myenumi}
\item[\textnormal{(i)}\hspace{2.6ex}] \textit{turbulent}, i.e. inherently three-dimensional and unsteady, possessing a range of dynamically active scales, and in which momentum and scalar diffusion occurs primarily through macroscropic fluctuations (e.g.  Reynolds stresses for momentum). This is quantified by a large Reynolds number (to be defined in \S~\ref{sec:methodology}, typically $Re \gg 10^3$) reflecting the overwhelming importance of inertial forces over viscous forces;

\item[\textnormal{(ii)}\hspace{2ex}] \textit{strongly-stratified}, i.e. flows in which the stable density stratification (typically in the form of a relatively sharp density interface) plays a significant role, for example through interfacial waves and the energetic cost of mixing the active scalar. This is quantified by a `relatively large' bulk Richardson number (to be defined in \S~\ref{sec:methodology}, typically $Ri_b = O(0.1-1)$), reflecting the non-negligible ratio of potential to kinetic energy in the system;

\item[\textnormal{(iii)}\hspace{1.5ex}] 
\textit{shear-driven}, i.e. flows in which turbulent kinetic energy is primarily extracted from a large-scale, largely-parallel, mean shear flow, away from solid boundaries. This configuration is implicit in the definition of the bulk Richardson number mentioned above, and excludes stratified turbulence forced by moving boundaries, internal waves, and other forms of spectral forcing (common in simulations using periodic geometry).

\item[\textnormal{(iv)}\hspace{1.5ex}] \textit{continuously-forced}, i.e. flow in which a continuous, steady flux of energy and unmixed fluid  respectively balance the turbulent dissipation and irreversible mixing. This allows a statistically-steady state of vigorous turbulence to be sustained for long periods of time (e.g.  $\gg 10^2$ advective time units), as in many flows of geophysical interest (excluding horizontal gravity currents which are inherently transient).

\end{myenumi}

As is often implicit in most of the geophysically-oriented literature on continuously-forced, shear-driven, and strongly-stratified turbulence, we further reduce the scope of this paper to flows that are: 

\begin{myenumi}
\item[\textnormal{(v)}\hspace{2.2ex}]
\textit{Boussinesq}, i.e. in which density differences are small enough (typically $\ll 5\%$ of the mean density) that they only play a relevant role through the acceleration of the reduced gravity;

\item[\textnormal{(vi)}\hspace{1.5ex}] \textit{high Prandtl number}, i.e. in which the ratio of momentum to scalar diffusion $Pr \equiv \nu/\kappa$ (also called the Schmidt number) is typically large, as is the case of temperature and salt stratification in water (where $Pr=7$ and $700$ respectively).
%, but less so to temperature stratification in air (where $Pr=0.7$) and excludes  most astrophysical scenarios where $Pr \ll 1$. 
As a result, the region of mean shear in which the mean-to-turbulent kinetic energy transfer occurs --~commonly referred to as the shear layer~-- is typically thicker than the density interface and embeds it (i.e. the ratio of shear layer to density interface thickness is $R >1$);

\item[\textnormal{(vii)}\hspace{1.2ex}] \textit{nearly-horizontal}, i.e. in which the normal to the mean shear flow and density interface is inclined with respect to the direction of gravity at most by a small angle (e.g.  $\theta < 10^\circ$), such that the main dynamics are horizontal (thus excluding plumes and exchange flows on steep slopes).

\end{myenumi}

We will derive insights from recently-available, three-dimensional velocity and density experimental data on exchange flows satisfying the above conditions, and exhibiting a variety of representative flow regimes (from laminar, to wavy, to intermittently turbulent and fully turbulent).
%, which give the time-resolved, simultaneous three-component velocity  and density fields in a three-dimensional volume.

The remainder of the paper is organised as follows. We motivate this study and explain our approach in \S~\ref{sec:context}, and introduce our methodology (experiment, notation and data sets) in  \S~\ref{sec:methodology}. %We then devote one section to each set of questions that we propose to make progress on: 
We will then make progress on the following sets of questions, to each of which we devote a section: 
\begin{myenumi}

\item[\textnormal{\S~4}\hspace{1.8ex}] What are the key non-dimensional parameters  ($Re,Ri_b,R$), the mean profiles, the forcing and dissipative mechanisms characterising these flows in various regimes? How do these compare with similar flows other observational, experimental, and numerical studies?
\item[\textnormal{\S~5}\hspace{1.8ex}] What is the distribution of the gradient Richardson number -- a key non-dimensional measure of the flow stability -- in various regimes? Does vigorous turbulence tends  `self-organisation', i.e. a kind of self-sustaining weakly-stratified equilibrium observed in other studies?
\item[\textnormal{\S~6}\hspace{1.8ex}] How to measure quantitatively  and characterise the character of intermittent or sustained turbulence using the concept of turbulent fraction with simultaneous velocity and density data? How do various data sets, spanning a range of non-dimensional parameters, compare and why?
\end{myenumi} 
Finally, we conclude in \S~7 and distill the key insights gained for the modelling of continuously-forced, shear-driven, stratified turbulence. 
%We tackle mean flows in \S~??, turbulent fractions in \S~??, and gradient Richardson number in \S~??. Finally, we conclude in \S~??.
In the companion Part 2 paper, we tackle the energetics, anisotropy, and parameterisation challenges.

\section{Context} \label{sec:context}

To provide context and motivation for our study, we discuss relevant field observations, numerical simulations, and laboratory experiments in  \S\S~\ref{sec:context-obs}-\ref{sec:context-exp} (for a summary table of the most recent and data-rich studies, see Appendix~\ref{sec:Appendix-litt-rev}). We then show where our study fits in and explain our approach in \S~\ref{sec:context-approach}.

\subsection{Field observations} \label{sec:context-obs}

Over the past decades, field observations have provided much data and insight on a variety of geophysical shear-driven stratified turbulent flows.

River plumes are outflow of buoyant water into the coastal ocean primarily forced by freshwater runoff \citep{mcpherson_turbulent_2019} and/or wind \citep{yoshida_mixing_1998}. The strength and spatial heterogeneity of turbulent mixing between these two water masses impact the physical, chemical, and biological properties of the developing coastal current \citep{macdonald_heterogeneity_2013}. 

Exchange flows between reservoirs of fluids at different densities are also highly relevant and occur on a variety of scales. At small scales, \cite{lawrence_summer_2004} investigated the exchange flow through a shallow ship canal connecting a small harbour to a lake undergoing seasonal, wind-driven cool upwelling, and the effects of this exchange on lake-shore pollution.  At larger scales, the strongly-stratified exchange flows in estuaries are primarily forced by periodic tides 
%(which over thousands of advective time units may still be considered quasi-steady)  
\citep{geyer_shear_1987,peters_microstructure_2000,macdonald_temporal_2008,tedford_observation_2009,geyer_mixing_2010}. At even larger scales, the relatively-steady baroclinic exchange flows through straits are weakly modulated by tides and influenced by the Earth's rotation, such as the much-studied strait of Gibraltar  \citep{farmer_flow_1988,armi_flow_1988,wesson_mixing_1994,macias_tidal_2006}.

In the deep Atlantic ocean, sill overflows of cold, Antarctic bottom water (AABW) through fractures such as the Romanche Trench are responsible for significant transport and mixing across ocean basins  \citep{ferron_mixing_1998,van_haren_extremely_2014}. In the upper-equatorial Pacific Ocean, deep periodic turbulent mixing events are caused by the interaction of a sustained vertical shear (between the wind-driven surface current and the opposing deep equatorial under-current) with a stable stratification modulated by diurnal solar heating \citep{smyth_narrowband_2011,smyth_diurnal_2013,smyth_pulsating_2017}.

Although field observations yield the most `realistic' data that one can hope for, they come at the cost of a limited control over the flow parameters, of great complexity in geometry and forcing conditions (wind, sun, tides, buoyancy, rotation), and of limited measurement abilities, all of which add up to make their general understanding challenging.

\subsection{Numerical simulations} \label{sec:context-dns}

A complementary approach is to isolate  physical mechanisms  by controlling the flow parameters, geometry, and forcing conditions in direct numerical simulations (DNSs) of the three-dimensional governing equations.

One of the key idealised model is the `stratified shear layer', or unforced parallel shear flow with hyperbolic tangent profiles for the velocity $\uu = (u(z),0,0)$ and density $\rho(z)$, free slip in the vertical direction and periodicity in the streamwise directions (see e.g.  \cite{smyth_anisotropy_2000}).  
Such mixing layers are prone to a range of linear instabilities, even in the presence of a single density interface, in particular the Kelvin-Helmholtz instability, whose initially two-dimensional billows undergo a zoo of secondary three-dimensional instabilities mediating the transition to turbulence at $Re=O(10^3)$ \citep{caulfield_anatomy_2000,mashayek_zoo_2012,mashayek_zoo_2012_2,mashayek_shear-induced_2013}. Mixing layers have complicated turbulent and mixing properties dependent on parameters such as the Reynolds, bulk Richardson and Prandtl numbers  \citep{salehipour_turbulent_2015,salehipour_diapycnal_2015,salehipour_turbulent_2016,salehipour_self_2018,watanabe_turbulent_2017}. The lack of forcing in these studies means that the turbulence is (at best) quasi-steady during a relatively short time before the initial kinetic energy is dissipated.

More recent studies focused on continuously-forced turbulence, using boundary forcing in the stratified plane Couette flow  \citep{zhou_diapycnal_2017,zhou_self-similar_2017}, and using a relaxation of the mean profiles to initial conditions in the stratified shear layer flow \citep{smith_turbulence_2021}. %\AL{Talk about `forced' DNS of  \cite{portwood_asymptotic_2019}. Maybe add them to table too.}

Although these studies provided exceptionally detailed quantitative insight, it remains challenging to perform continuously-forced simulations of flows satisfying all criteria in \S~1.1 with parameters relevant to field observations (in particular the Reynolds and Prandtl numbers, see table~1). More fundamentally, simulations are approximations of idealised equations, which typically assume (among others): no rotation, incompressibility, the Boussinesq approximation, a linear equation of state, a single active scalar, spatially homogenenous momentum and scalar diffusivity with idealised values of $Pr$, as well as simplistic geometry, initial conditions, boundary conditions and forcing. The sensitivity of the governing equations to such real world imperfections (e.g.  through singular perturbations) remains an open question, and as such, we ought to remain critical of the relevance of numerical simulations to explain field observations.

\subsection{Laboratory experiments}  \label{sec:context-exp}

Laboratory experiments offer a valuable intermediate approach, by allowing more control over flow parameters, geometry, forcing and measurements than in  the field, while retaining some of the inherent complexity of `real' flows discarded in simulations.  

A few laboratory flows satisfying all seven criteria in \S~1.1 have been studied (see Appendix~\ref{sec:Appendix-litt-rev}), typically using a combination of qualitative flow visualisations and quantitative single-plane velocity/density data at relatively low resolution in space and/or time. \cite{strang_entrainment_2001} studied the entrainment at a weakly-turbulent interface in a closed-loop recirculating flume driven by disk pumps (known as the Kovasznay flume after \cite{odell_new_1971}). %their $Ri_s$ based on shear layer is our $Ri_b$ (not their $Ri_b$) hence the Rib I quote take this into account
%ignore early experimental studies that don't have any 'real' turbulent data (any sort of fluctuation etc) and are mostly qualitative such as \citep{koop_instability_1979,}
\cite{odier_fluid_2009,odier_entrainment_2014,odier_stability_2017}  studied the similar problem of entrainment and mixing of a turbulent wall jet developing into a sloping gravity current over a dense quiescent layer.  \cite{meyer_stratified_2014,lefauve_buoyancy_2020} studied the transitions between flow regimes in exchange flows taking place in an inclined duct.

The added value of laboratory experiments such as those cited above in the three-pronged (observational, numerical, experimental) approach has so far been somewhat limited by the challenge of obtaining high-resolution, three-dimensional measurements of turbulent flow fields. 

However, such measurements are now becoming available. The novel scanning stereo PIV-PLIF system introduced in \cite{partridge_versatile_2019} achieves simultaneous measurements of the density and three-component velocity fields in a three-dimensional volume. Using this novel system in the stratified inclined duct geometry, \cite{lefauve_structure_2018} studied the three-dimensional stability properties of interfacial Holmboe waves, and \cite{lefauve_regime_2019} studied the time- and volume-averaged energy budgets of 16 data sets spanning a range of flows on either side of the turbulent transition.

\subsection{Approach} \label{sec:context-approach}

To achieve the objectives set out in \S~\ref{sec:intro}, we will further analyse the 16 experimental data sets of \cite{lefauve_regime_2019}. These cutting-edge density and velocity data are ideally suited for our purpose since they are non-intrusive, three-dimensional and three-component, simultaneous, high-resolution (in space and time), and accurate.

The key methodological differences between this paper and  \cite{lefauve_regime_2019} are: (i) our focus, in this paper, on turbulent fluctuations and statistics inside the shear layer (as opposed to volume-averages including wall effects); (ii) our analysis of these data in a framework consistent with the observational and numerical literature on stratified shear layers (in particular the non-dimensional notation), allowing for more direct comparison and added value to the general community. We introduce this methodology in the next section.

\section{Methodology} \label{sec:methodology}

We introduce our experimental set-up in \S~\ref{sec:setup}, the hydraulic non-dimensionalisation of variables in \S~\ref{sec:hydraulic-nondim}, and the non-dimensional rescaling of experimental data suited to comparison with canonical stratified shear layers in \S\S~\ref{sec:shear-layer-rescaling}-\ref{sec:canonical-SL}. Finally, we introduce our data sets in \S~\ref{sec:datasets}.

\subsection{The stratified inclined duct experiment} \label{sec:setup}

\begin{figure}
\centering
\includegraphics[width=\textwidth]{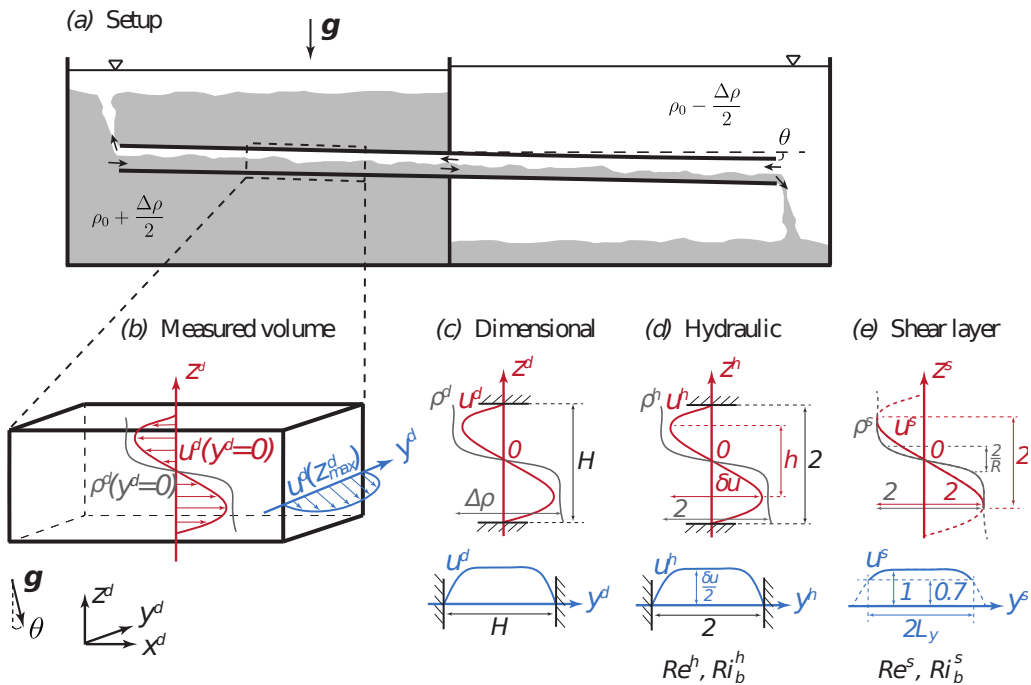}
    \caption{Set-up and notation. \emph{(a)} The stratified inclined duct (SID) experiment (see \S~\ref{sec:setup}). \emph{(b)} Measured duct sub-volume,  dimensional coordinate system $(x^d,y^d,z^d)$ and flow variables $\rho^d, u^d$, with three key schematic flow profiles (in grey, red, and blue) at $y^d=0$ and $z^d_{max}$. We then compare these key flow profiles \emph{(c)} in dimensional units; \emph{(d)} after the hydraulic non-dimensionalisation in \eqref{def-hyd-nondim};  \emph{(e)} after the shear layer rescaling in \eqref{def-shear-rescaling}, yielding profiles comparable to canonical $\tanh$ shear layers (we discard the dashed line profiles outside the main shear layer $|y^s|>L_y, |z^s|>1$).}
    \label{fig:setup}
\end{figure}

We consider the stratified inclined duct (SID) experiment sketched in figure~\ref{fig:setup}\emph{(a)}. We study the steady-state exchange flow sustained inside a long ($L=1350$~mm) duct of square cross-section ($H=45$~mm), inclined at a small angle $\theta$, connecting two large reservoirs initially filled with aqueous salt solutions ($Pr=700$) of different densities $\rho_0 \pm \Delta \rho/2$. This exchange flow naturally achieves continuously-forced, shear-driven, strongly-stratified turbulence at the interface, i.e. away from the solid duct boundaries (a good approximation to free shear).

The SID experiment has been studied in detail in prior publications, and we refer the reader to these for further details about the set-up: \cite{meyer_stratified_2014} (hereafter ML14, see their \S~2),   \cite{lefauve_structure_2018} (hereafter LPZCDL18, see their \S~3), \cite{lefauve_regime_2019} (hereafter LPL19, see their \S~1-2), and \cite{lefauve_buoyancy_2020} (hereafter LL20, see their \S~2).

\subsection{Hydraulic non-dimensionalisation} \label{sec:hydraulic-nondim}

Like all exchange flows, we expect the flow in the SID to be forced by a mean streamwise pressure gradient of opposite directions in each layer, even when the duct is horizontal. This streamwise pressure gradient results from the expectation that the pressure is constant along the plane of neutral density $\rho=\rho_0$ and that the ends of the duct sit in reservoirs of different densities (see \citealp{lefauve_waves_2018}, \S~1.2.2). The resulting two-layer hydraulic flow has (dimensional) peak-to-peak velocity jump set (approximately) by the buoyancy velocity scale $\Delta U \equiv 2\sqrt{g'H}$, where $g'\equiv g\Delta \rho/\rho_0$ is the reduced gravity.

As commonly done is the hydraulic community, LPZCDL18, LPL19 and LL20 used halves of the total density difference ($\Delta \rho/2$), duct height ($H/2$), and velocity jump ($\Delta U/2$) to non-dimensionalise all variables: 
\begin{equation}  \label{def-hyd-nondim}
\quad \rho^h \equiv \frac{\rho^d-\rho_0}{\Delta\rho/2}, \quad  \uu^h = \frac{\uu^d}{\Delta U/2}, \quad \xx^h \equiv \frac{\xx^d}{H/2}, \quad t^h \equiv \frac{t^d}{H/\Delta U},
\end{equation}
where $\rho$ and $\uu=(u,v,w)$ are the density and velocity fields respectively, $\xx=(x,y,z)$ is the position vector in the coordinate system defined in figure~\ref{fig:setup}\emph{(b)}, and $t$ is the time. The superscripts ${}^d$ and ${}^h$ denote, respectively, dimensional and non-dimensional hydraulic variables.

This hydraulic non-dimensionalisation leads to the natural definitions of `input' Reynolds and bulk Richardson numbers (i.e. depending only on parameters set by the experimentalist) that we refer to in this paper as `hydraulic' Reynolds number and `hydraulic' bulk Richardson numbers:
\begin{equation} \label{def-Reh-Ribh}
    Re^h \equiv \frac{\dfrac{
    \Delta U}{2}\dfrac{H}{2}}{\nu} = \frac{\sqrt{g'H}H}{2\nu} = 1.42\times 10^4 \sqrt{\frac{\Delta \rho}{\rho_0}}, \qquad  Ri^h_b \equiv \frac{\dfrac{g}{\rho_0}\dfrac{\Delta \rho}{2}\dfrac{H}{2}}{\Big(\dfrac{\Delta U}{2}\Big)^2} = \frac{1}{4}.
\end{equation}

The flow in the SID is not only forced by a streamwise pressure gradient, but also by the projection of gravity $\mathbf{g}$ along $x$ due to the tilt angle $\theta>0$ of the duct (in this paper between $1^\circ$ and $6^\circ$, as sketched in figure~\ref{fig:setup}\emph{(a)}). These two forcing mechanisms yield a variety of flow regimes: from laminar flow with flat interface ($\LL$ regime), to mostly-laminar Holmboe waves propagating at the interface ($\HH$ regime) to intermittent turbulent ($\II$ regime) to fully-developed turbulence ($\TT$ regime). These flow regimes and their transitions have been mapped in the ($\theta,Re^h$) plane and discussed extensively in ML14, LPL19 and LL20. 

One of the key conclusions of these past studies of the SID experiment is that, while the dimensional peak-to-peak velocity scale of $u^d$ is primarily set as $\Delta U \equiv 2\sqrt{g'H}$ by the longitudinal pressure gradient (hydraulic scaling), the actual (measured) non-dimensional peak-to-peak magnitude of $u^h$ (in an $x$- and $t$- averaged sense) is a complicated $O(1)$ function of $Re_h$ and $\theta$. 

To illustrate this point, we define three key profiles in the duct sub-volume of figure~\ref{fig:setup}\emph{(b)}: the vertical profiles of density $\rho^d$ (in grey) and streamwise velocity $u^d$ (in red) in the vertical plane of maximum velocity (the mid-plane $y^d=0$), as well as the spanwise profile of $u^d$ (in blue) in the horizontal plane of maximum velocity ($z^d=z^d_{max}$). These three profiles are drawn schematically in dimensional variables in figure~\ref{fig:setup}\emph{(c)} and after the hydraulic non-dimensionalisation \eqref{def-hyd-nondim} in figure~\ref{fig:setup}\emph{(d)}. 

Figure~\ref{fig:setup}\emph{(d)} shows that while the duct height, duct width, and the magnitude of the total density jump are always $2$, the peak-to-peak velocity $\delta u$ and the height between the velocity peaks $h$ are both \emph{a priori} unknown.

\subsection{Shear layer rescaling} \label{sec:shear-layer-rescaling}

In order to analyse our data in a non-dimensional framework quantitatively consistent with most of the literature on stratified shear layers, we define the following shear layer rescaling, using halves of the `output' (measured) velocity jump $\delta u/2$ and shear layer depth $h/2$:
\begin{equation} \label{def-shear-rescaling}
\quad \rho^s \equiv  \rho^h, \quad  \uu^s = \frac{\uu^h}{\delta u/2}, \quad \xx^s \equiv \frac{\xx^h}{h/2}, \quad t^s \equiv \frac{t^h}{h/\delta u},
\end{equation}
where the superscripts $^h$ and  $^s$ denote respectively the hydraulic non-dimensional variables defined in \eqref{def-hyd-nondim} and the new shear layer variables.

Figure~\ref{fig:setup}\emph{(e)} shows that the rescaled total velocity jump and shear layer depth are now always 2. Since the symmetry of the flow with respect to $y^s, z^s=0$ is sometimes approximate, we further shift the $y^s,z^s$ axes to centre them such that the bounds of the shear layer are exactly $|y^s| \le L_y, |z^s| \le 1$. The total shear layer width $2L_y$ is the smallest width in which both profiles $u^s(y^s,z^s=\pm 1)$ are at least $70~\%$ of their extremum value (typically $2L_y \approx 3$ in our data). 

We also define the velocity-to-density thickness ratio $R$, where $2/R$ is the typical density layer thickness defined as spacing between the points at which $\rho^s = \pm \tanh(1) \approx 0.76$ (giving typically $R>1$ when $Pr\gg 1$). 

The dashed lines in figure~\ref{fig:setup}\emph{(e)} denote flow outside the shear layer, where velocities decay to zero to satisfy the no-slip boundary condition at the four duct walls. In the remainder of this paper we ignore wall effects by discarding data outside the shear layer.

This rescaling leads to the definitions of the following `shear' Reynolds number and `shear' bulk Richardson number:
\begin{equation}\label{def-Res-Ribs}
Re^s \equiv Re^h  \, \frac{\delta u \, h}{4}, \qquad Ri_b^s \equiv Ri_b^h \, \dfrac{(h/2)}{(\delta u/2)^2} \equiv  \dfrac{h}{2(\delta u)^2}.
\end{equation}
Note that our $Ri_b^s$ is sometimes called $Ri_0$ or $J$ in the literature.

In the remainder of this paper, unless specified otherwise, we use the shear layer variables defined in \eqref{def-shear-rescaling} and drop the superscript $^s$ (except in $Re^s$ and $Ri_b^s$, for clarity). 

The corresponding governing Navier-Stokes equations in shear layer variables under the Boussinesq approximation are then
\begin{subeqnarray}\label{eq-motion}
\bnabla \cdot \uu &=& 0, \slabel{eq-motion-1}\\ 
\frac{\p \uu}{\p t}  + \uu \cdot \bnabla \uu &=& -\bnabla p   + Ri_b^s \, \rho \, (-\mathbf{\hat{z}}  +  \sin \theta \, \mathbf{\hat{x}}) + \frac{1}{Re^s} \bnabla^{2}\uu
, \slabel{eq-motion-2}\\
\frac{\p \rho}{\p t} + \uu \cdot \bnabla \rho &=& \frac{1}{Re^s \, Pr} \bnabla^{2} \rho, \slabel{eq-motion-3}
\end{subeqnarray}
where we assumed that $\cos \theta \approx 1$ in nearly-horizontal flows (accurate to better than $1~\%$ in this paper). We discuss boundary conditions next.

\subsection{Comparison with canonical shear layers} \label{sec:canonical-SL}

The above rescaling makes our data (figure~\ref{fig:setup}\emph{(e)}), non-dimensional parameters \eqref{def-Res-Ribs} and governing equations \eqref{eq-motion} comparable to those found in studies of canonical stratified shear layers defined by the initial ($t=0$) profiles:
\begin{equation} \label{canonical-shear-layer}
u(z)=-\tanh(z), \qquad \rho(z)= -\tanh(R z).
\end{equation}
Note the minus signs, typically absent in the literature, but retained here for historical reasons and of minor significance (note that some studies prefer to use the buoyancy field, here simply equal to $-\rho$). A relatively small number of studies opt for a non-dimensionalisation based on the total (as opposed to half) velocity jump $\delta u$ and shear layer depth $h$, making their Reynolds number four times as large as ours, and their bulk Richardson number half as large as ours. The values of $Re^s$ and $Ri_b^s$ in Appendix~\ref{sec:Appendix-litt-rev} have been estimated and/or converted from various studies to match our definitions consistent with the governing equations \eqref{eq-motion} and the canonical $\tanh$ model \eqref{canonical-shear-layer}.

We however note at least five interesting differences between our rescaled SID flows and most canonical $\tanh$ shear layers:

\begin{myenumi}
\item[\textnormal{(i)}\hspace{2.8ex}]  our rescaled profiles in figure~\ref{fig:setup}\emph{(e)} are understood as `mean flows' averaged in the horizontal direction and over a long-time equilibrium, as opposed to carefully designed initial conditions; 
\item[\textnormal{(ii)}\hspace{2.1ex}]  our velocity at the top and bottom boundaries of the shear layer reaches approximately $\pm 1$ (in the mid-plane $y=0$) and $\pm 0.8-0.9$ (when averaged in $y$ across the layer), as opposed to the more modest $\tanh(1)\approx \pm 0.76$; 
\item[\textnormal{(iii)}\hspace{1.4ex}] our vertical shear at the top and bottom boundaries of the shear layer is zero $\partial_z u (z \approx \pm 1)=0$, because of the influence of the nearby top and bottom walls, as opposed to the typical free-slip boundary conditions at $z\rightarrow \pm \infty$;  
\item[\textnormal{(iv)}\hspace{1.6ex}] our spanwise velocity gradient is non-zero at the spanwise edges of the shear layer $\partial_y u (y = \pm L_y)\neq 0$, because of the influence of the nearby side walls, as opposed to the typical periodic boundary conditions in $y$;  
\item[\textnormal{(v)}\hspace{2.1ex}] our long-time equilibrium is achieved by a gravitational body force along $x$ ($Ri_b^s \sin \theta \rho$) and by non-periodic boundary conditions along $x$ responsible for both a mean horizontal pressure gradient and a mean horizontal buoyancy flux (continuously replacing partially mixed fluid in the duct by unmixed fluid from the reservoirs), all of which are typically absent in canonical shear layer simulations. 
\end{myenumi}

\subsection{Data sets} \label{sec:datasets}

We use 16 sets of time-resolved, volumetric data of the density and three-dimensional, three-component velocity (3D-3C) $(u,v,w,\rho)(x,y,z,t)$ freely available online \cite{lefauve_research_2019}. These were obtained by successive $x-z$ planar measurements of stereo particle image velocity (sPIV) and laser induced fluorescence (LIF) performed simultaneously in a rapid, continuous, back-and-forth scanning motion across $y$ to reconstruct successive three-dimensional volumes. In all experiments, the duct streamwise aspect ratio was $30$, the duct spanwise aspect ratio was $1$ (square), and the Prandtl number was $Pr \approx 700$ (NaNO$_3$/NaCl salt solutions with matched refractive indices).  For more information on the set-up, scanning technique, and post-processing (including imposing $\bnabla \cdot \uu =0$ in all volumes), we refer the reader to  \cite{partridge_versatile_2019} (their \S~3-4),  LPZCDL18 (their \S~3.3-3.4) and LPL19 (their \S~3.1-3.2).

To suit the objectives of the present paper, these data sets were modified in the following two ways. First, as explained in \S~\ref{sec:shear-layer-rescaling}, we only retain data in the shear layer (by discarding the near-wall data dashed in figure~\ref{fig:setup}\emph{(e)}). The final size of each volume  $(2L_x,2L_y,2L_z)$  is given in Appendix~\ref{sec:Appendix-data} (in shear layer units, where by definition $2L_z=2$), together with the total remaining number of grid points in each direction $(n_x,n_y,n_z)$, and the resulting resolution  $(\Delta x, \Delta y, \Delta z) \equiv (2L_x/n_x, 2L_y/n_y, 2/n_z)$). Second, small errors in the initial levels of free surfaces in each reservoir (figure~\ref{fig:setup}\emph{(a)}) caused small barotropic (net) flow oscillations between the two reservoirs, which decayed exponentially with time. Data sets showing these early-time damped oscillations were cropped in time to keep only the later-time, statistically-steady part of the flow. The resulting length of each data set $L_t$ (in shear-layer advective time units) is given in Appendix~\ref{sec:Appendix-data} together with the total number of volumes $n_t$ and the temporal resolution $\Delta t\equiv L_t/n_t$ (or time taken to scan a volume from wall to wall i.e. $y^h=\pm 1$).

\begin{table}
  \begin{center}
\def~{\hphantom{0}}
\setlength{\tabcolsep}{6pt}
  \begin{tabular}{lccccccc}
      Name &    \multicolumn{2}{c}{Input params.}   &  \multicolumn{5}{c}{Output params.} \\   \cmidrule(l{3pt}r{3pt}){2-3} \cmidrule(l{3pt}r{3pt}){4-8} 
             &  $\theta$ & $Re^h$   & $\delta u$ & $h$ & $R$ & $Re^s$ & $Ri_b^s$   \\ [5pt]
L1 & 2 & 398  & 0.794 & 1.03 & 13.0 & 81 & 0.812  \\[5pt] % m141
H1 & 1 & 1455 & 0.973 & 1.08 & 8.9 &  381 & 0.567   \\ % m135
H2 & 5 & 402  & 2.00 & 1.01 & 7.2  &   204 & 0.127  \\ %m125
H3 & 2 & 1059 & 1.43 & 1.11  & 11.3  &   422 & 0.273\\ % m129
H4 & 5 & 438  & 1.85 & 1.00 & 10.0    &   203 & 0.146   \\[5pt] % m133
I1 & 2 & 1466 & 1.60 & 0.907 & 5.8 &  531 & 0.178  \\ % m130
I2 & 2 & 1796 & 1.70 & 1.14 &  5.3 & 872 & 0.196  \\ % m131
I3 & 2 & 2024 & 1.65 & 1.06 & 4.4 &   891 & 0.194 \\ % m132
I4 & 6 & 777  & 2.56 & 1.30 & 2.4 &   646 & 0.099  \\ % m144
I5 & 5 & 956  & 2.02 & 1.26  & 2.3 &  607  & 0.155   \\ % m126
I6 & 6 & 798  & 1.93 & 1.29 & 2.2 &   497 & 0.173 \\ % m145
I7 & 3 & 1580 & 1.91 & 1.20  & 2.7 &   905 & 0.163  \\ % m146
I8 & 5 & 970  & 2.31 & 1.26 & 2.4  &   708 & 0.118 \\[5pt] % m138
T1 & 3 & 2331 & 1.93 & 1.31 & 2.1 &   1479 & 0.176  \\ % m147
T2 & 6 & 1256 & 2.30 & 1.42 & 1.8  &   1030 & 0.134 \\ % m143
T3 & 5 & 1516 & 2.17 & 1.39 & 1.9 &   1145 & 0.147   \\ % m134
  \end{tabular}
  \caption{List of the 16 volumetric data sets used, with input parameters $\theta$ and $Re^h$ (note $Ri_b^h = 1/4$), sorted by increasing $\theta Re^h$, and thus, by flow regime $\LL$, $\HH$, $\II$, $\TT$ (as in LPL19 table~2). The output parameters follow the shear layer rescaling in \S~\ref{sec:shear-layer-rescaling} and figure~\ref{fig:setup}\emph{(e)}. The four parameters $(\theta,Re^s,Ri^s_b,R)$ are necessary and sufficient to describe the model in \eqref{eq-motion}-\eqref{canonical-shear-layer}.}
\label{tab:dataset}
  \end{center}
\end{table}

The key properties of our 16 data sets are shown in table~\ref{tab:dataset}. One flow belongs to the laminar ($\LL$) regime (named L1), four flows to the Holmboe wave ($\HH$) regime (named H1-H4), eight flows to the intermittently turbulent  ($\II$) regime (named I1-I8) and three flows to the fully turbulent ($\TT$) regime (named T1-T3).  These data sets are ordered by increasing values of the product of input parameters $\theta Re^h$, as in LPL19 (see their table~2) who showed that $\theta Re^h$ controlled the time- and volume-averaged kinetic energy dissipation and thus the transitions between flow regimes (note $\sin \theta \approx \theta$ in our nearly-horizontal flows). The output parameters $\delta u, h, R, Re^s$ and $Ri^b_s$ were determined as explained in \S~\ref{sec:shear-layer-rescaling}, where the key profiles drawn in figure~\ref{fig:setup} were interpreted as $x$- and $t$- averages over the data set (i.e. over $x\in [0,2L_x]$, and $t\in [0,L_t]$). We discuss the values of these output parameters next.

%%%%%%%%%%%%%%%%%%%%%%%%%%%%%%%%%%%%%%%%% 

\section{Flow parameters and Reynolds averages} \label{sec:mean_prop}

In this section we further characterise our data sets with three key pieces of information: the output flow parameters in \S~\ref{sec:output_params}, the mean flow profiles in \S~\ref{sec:mean_flows}, and the Reynolds-averaged balances sustaining these mean flows in \S~\ref{sec:reynolds-avg}.

\subsection{Output parameters} \label{sec:output_params}

\begin{figure}
\centering
\includegraphics[width=1.03\textwidth]{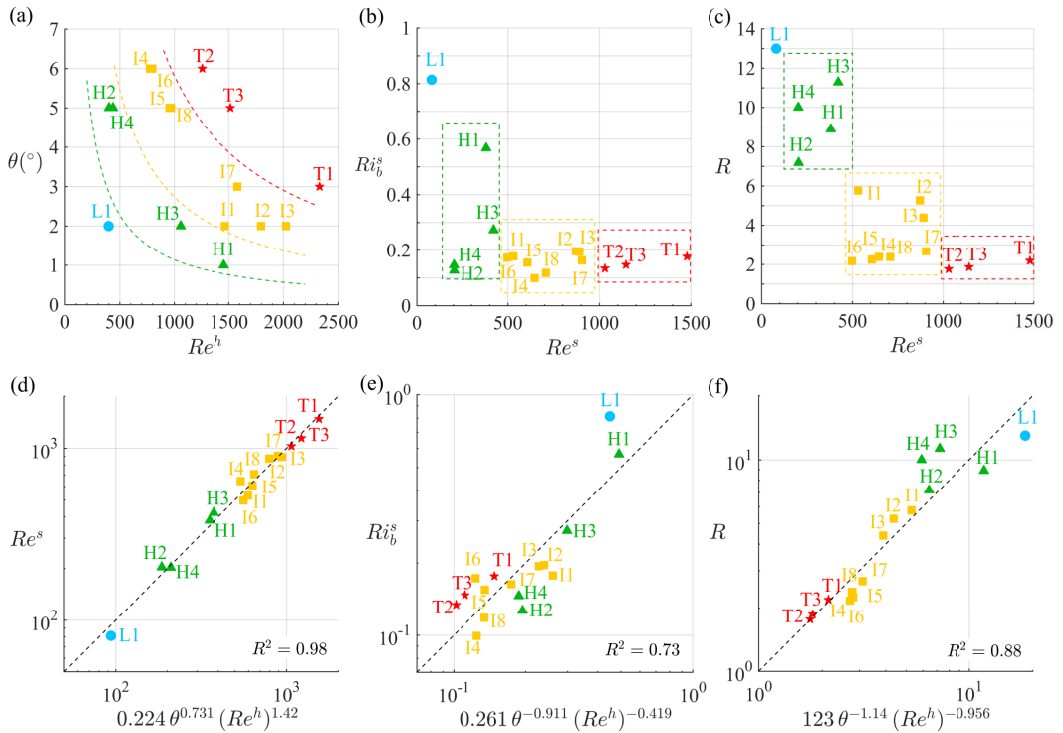}
    \caption{Maps of all 16 data sets of table~\ref{tab:dataset} in the space of \emph{(a)} input parameters $(Re^h,\theta)$ and \emph{(b-c)} output parameters $(Re^s,Ri^s_b,R)$. The dashed curves in \emph{(a)} are the regime transitions in $\theta Re^h =$~const. previously given by LPL19. The dashed rectangles in  \emph{(b-c)} highlight the fact that regimes also occupy distinct regions in the output space. \emph{(d-f)} Best power law fit (least-squares linear regression in log-log space) of the output parameters $(Re^s,Ri^s_b,R)$ by the input parameters $\theta,Re^h$ (fit \emph{vs} actual value, the dashed line denoting equality). In power law scalings, $\theta$ is always expressed in radians.}
    \label{fig:output_params}
\end{figure}

In figure~\ref{fig:output_params} we plot maps of all 16 data sets of table~\ref{tab:dataset} in the space of input parameters $(Re^h,\theta)$ (panel~\emph{a}) and in the space of our three independent output parameters: $(Re^s,Ri^s_b,R)$ (panels~\emph{b-c}). We also show the power law regressions of the output parameters with respect to the input parameters  (panels~\emph{d-f}).

First, we see that the bulk Richardson number $Ri_b^s$ and the velocity-to-density thickness ratio $R$ typically decrease as the Reynolds number $Re_s$ increases, and appear to reach asymptotic values in the turbulent regime (panels~\emph{b-c}). In other words, the relatively wide and uniformly-sampled region of the input space (panel~\emph{a}) is mapped by the mean flow dynamics into a relatively narrow and specific region of the output space (panels~\emph{b-c}). The flow dynamics also have an inherent degree of randomness making them not generally repeatable, because we see that near-identical input parameters can be mapped into fairly different output parameters (e.g.  compare the couples H2/H4, I4/I6, and I5/I8 in panel~\emph{a} and panels~\emph{b-c}). The above two observations mean that the experimentalist (or the numericist simulating these flows) has only a limited (and not fully understood yet) ability to control the output parameters from the input parameters. 

Second, we see that different qualitative flow regimes $\LL,\HH, \II, \TT$ (respectively in blue, green, yellow and red) occupy distinct and well-defined regions in the $(Re^s,Ri_b^s,R)$ output space (sketched in panels~\emph{b-c} by the dashed rectangles). This  result, which  implies that the different flow regimes reflect different physics, could not simply be predicted \emph{a priori} from the previously-known result that regimes occupy distinct regions in the $(Re^h,\theta)$ input space (sketched in panel~\emph{a} by the dashed curves of  LPL19). 

In the output space (panels~\emph{b-c}), the transition from stable laminar flow to regular Holmboe waves ($\LL \rightarrow \HH$) is correlated with $Re^s \gtrsim 100-200$, $Ri^s_b \lesssim 0.6-0.8$ and $R\lesssim 12$, values that are consistent with the triggering of Holmboe instability. The transition to intermittent turbulence ($\HH\rightarrow \II$) is correlated with $Re^s\gtrsim 500$, $Ri^s_b\lesssim0.2$ and $R \lesssim 7$, while the transition to sustained turbulence ($\II \rightarrow \TT$) is correlated with $Re^s \gtrsim 1000$, and the asymptotic values $Ri^s_b \approx 0.15$ and $R\approx 2$. %\pfl{\it I would delete the next sentence as it repeats the one in the previous paragraph - and I think it is, in general, wrong to tell the reader what is interesting. }These are further interesting results that add to our understanding of regime transitions and that could not simply be predicted from the previously-known result that regimes transitions are caused by threshold values of $\theta Re^h$ (LPL19).

Third, we observe that the maps in the output space are not entirely consistent with the use of $\theta Re^h$ as a proxy for flow regimes and as a means to quantitatively order flows within regimes (based on their closeness to another regime), as was done in LPL19 and in our nomenclature of the data sets.
%, where, e.g., I8 has a higher $\theta Re^h$ than I1 suggesting that I8 is close to T1 and I1 is close to H4). 
For example, we see in panels~\emph{b-c} that I2/I3 are closer to $\TT$ flows than I6/I8 are, and that T2/T3 are closer to $\II$ flow than T1 is, whereas our nomenclature suggests otherwise in both cases. We also see in  panel~\emph{b} that, although the five flows I4-I8 have near-identical $\theta Re^h$, they stretch all the way from the $\HH$ transition to the $\TT$ transition.

Fourth, we note that vigorous turbulence can be sustained even at relatively low $Re^s\sim 1000$ due to the continuously-forced nature of SID flows. Indeed, our largest value $Re^s \approx 1500$ in the $\TT$ regime is a factor three to four lower than the values of $4000-6000$ investigated in the latest numerical simulations of stratified shear layers  \citep{salehipour_diapycnal_2015,salehipour_turbulent_2016,smith_turbulence_2021}. Although much higher $Re^s \approx Re^h = O(10^4-10^5)$ can readily be achieved in the SID experiment (see LL20), they are not shown here because they remain out of reach of detailed quantitative measurements due to limitations in the spatio-temporal resolution of the scanning sPIV/LIF technique (discussed in LPL19, Appendix~A).

Fifth, we study the power law regression (best fit) of the output parameters with respect to the input parameters. The scaling $Re^s \propto \theta^{0.73} (Re^h)^{1.4}$ (panel~\emph{d}) is an excellent fit, since most symbols lie close to the dashed line (the coefficient of determination is $r^2=0.98$). This shows that $\theta$ plays a key role in setting the non-dimensional  scales $\delta u, h$, and thus $Re^s$ (remembering from \eqref{def-Res-Ribs} that $Re^s\equiv \delta u  \, h \, Re^h/4$). However $Ri_b^s \propto \theta^{-0.91} (Re^h)^{-0.42}$ (panel~\emph{e}) is a poorer fit ($r^2=0.73$); although $Ri_b^s$  tends to decrease with both $\theta$ and $Re^h$, the data have more variability than can be explained by a simple power law.  Finally, $R \propto \theta^{-1.14} (Re^h)^{-0.42}$ (panel~\emph{f})  is a good fit ($r^2=0.88$), showing that the non-dimensional density layer thickness $2/R$ tends to increase strongly with $\theta$, and more weakly with $Re^h$. This is consistent with the findings of LL20 (see their figures 7-8), who applied a similar (though higher-order) fitting to density layer thickness data obtained by shadowgraph image analysis in various duct geometries, across hundreds of experiments covering a wider range of $\theta,Re^h$ than in the present paper.

\subsection{Mean flows} \label{sec:mean_flows}

%To further characterise our data sets, 
We now turn to mean flows. Here, and in the remainder of this paper, we define the averages for any flow variable $\phi$ as follows:
\begin{subeqnarray}\label{def-avg}
\bar{\phi}(y,z) &\equiv& \langle \phi \rangle_{x,t} \equiv  \frac{1}{2L_t L_x} \int_0^{L_t} \int_0^{2L_x}\phi \ \d x\, \d t, \slabel{def-avg-1} \\
\langle \phi \rangle &\equiv& \langle \phi \rangle_{x,y,z,t} \equiv  \frac{1}{4L_y} \int_{-L_y}^{L_y} \int_{-1}^{-1}\bar{\phi} \ \d y\, \d z,\slabel{def-avg-2}
\end{subeqnarray} 
where $\langle \phi \rangle_{i}$ denotes averaging with respect to any coordinate $i$, $\bar{\phi}$ denotes specifically $x$- and $t$-averaging (what we usually call the `mean'), and $\langle \phi \rangle$ denotes time- and volume-averaging. All averaging is performed using accurate trapezoidal numerical integration.

\begin{figure}
\centering
\includegraphics[width=0.95\textwidth]{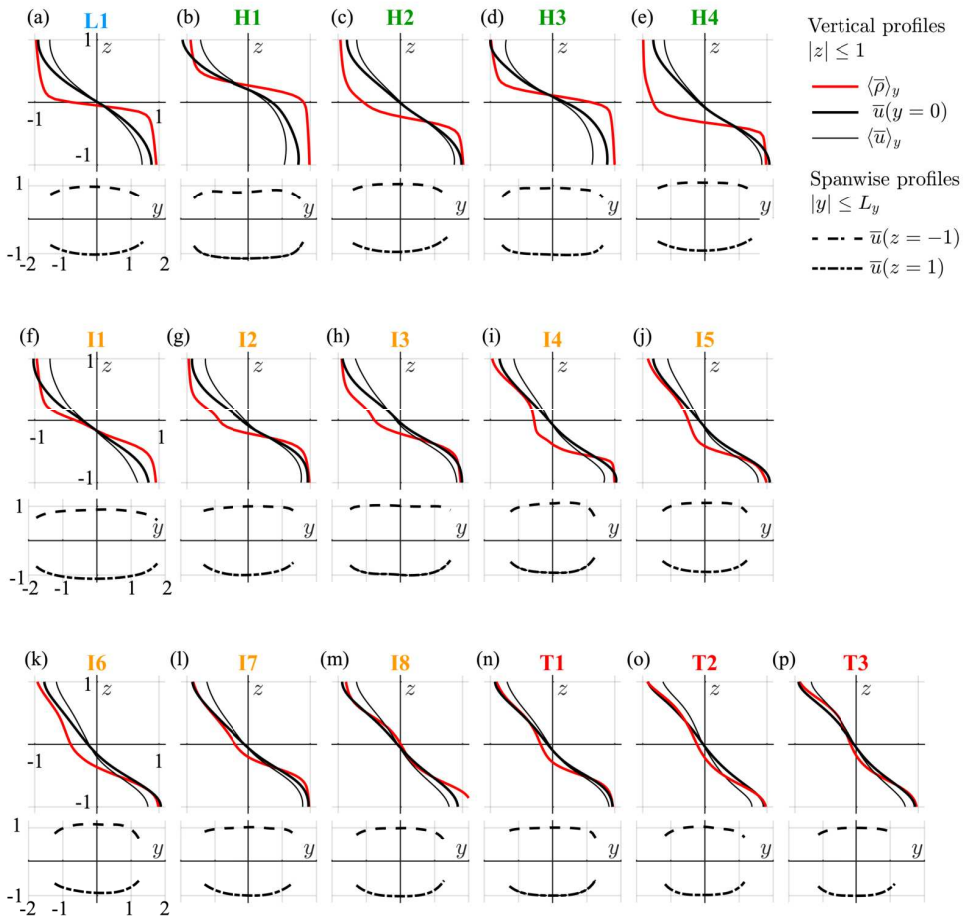}
    \caption{Mean flows in all 16 data sets of table~\ref{tab:dataset}: vertical profiles (top sub-panels of \emph{a-p}) and spanwise profiles (bottom sub-panels of \emph{a-p}), see legend for details. Legend and axes limits are identical in all respective sub-panels.}
    \label{fig:mean_flows}
\end{figure}

Figure~\ref{fig:mean_flows} shows the mean streamwise velocity $\bar{u}$ and density $\bar{\rho}$ from all 16 data sets. 
Each panel \emph{(a-p)} corresponds to a data set; the top sub-panels show vertical profiles (both the mid-plane velocity maximum $\bar{u}(y=0)$ and $y$-averages $\langle \bar{u}\rangle_{y}$, $\langle \bar{\rho}\rangle_{y}$,  across the whole shear layer $|y|\le L_y$), while the bottom sub-panels show the spanwise profiles at the top and bottom edges of the shear layer $\bar{u} (z=\pm 1)$.

First, we see that the horizontal profiles $\bar{u}(z=\pm 1)$ show excellent spanwise symmetry (about the $y=0$ plane, the `real' mid-plane of the duct), as expected from the symmetry of the duct. In the shear layer region plotted here ($|y|\le L_y$), where we recall that by definition velocities are at least 70~$\%$ of their extremum, we see a fairly extended flat region where $\p_y \bar{u} \approx 0$. This region  typically occupies at least $|y|\le 1$, and is slightly wider in some data sets, with no obvious dependence on flow parameters (not even on $Re^s$, surprisingly). This suggests that despite the existence of side walls in our experiment, our flows contain shear layers whose mean flows exhibit very little spanwise variations over an extent at least as large as the vertical extent ($|z|\le 1$). Closer to the spanwise edges of the shear layer, our mean flows have $\p_y \bar{u} \neq 0$ and the resulting effects of this spanwise shear on the turbulence can in principle be investigated (which is not possible in simulations with periodic boundary condition in $y$).

Second, we see in some data sets that the vertical profiles of $\bar{\rho}$ and $\bar{u}$ are `offset' with respect to one another, i.e. the $\bar{\rho}=0$ and $\bar{u}=0$ levels are not collocated and $\bar{\rho}\bar{u}<0$ (particularly visible in panels~\emph{c,e,g,h,i,j,k}). 

In the Holmboe wave regime, where the density interface is sharp ($R>7$) and $\tanh$-like, this offset gives rise to asymmetric (i.e. one-sided) Holmboe waves  (in H2 and H4). (For further empirical observations of this offset, see \cite{lefauve_waves_2018}~\S~3.2.2, and for visualisations and explanation of these waves in H4, see LPZDCL18.) By contrast, the absence of offset gives rise to symmetric (i.e. two-sided) Holmboe waves (in H1 and H3). (For a visualisation of these waves in H1, see LPL19 figure~3\emph{g-j}.) We note that this offset is inconsistent with the effects of gravitational forcing alone (see the term $Ri^s_b \sin\theta \rho$ in \eqref{eq-motion-2}) and, therefore, it suggests the important role of a horizontal pressure gradient with a more complicated $z$ profile than hitherto assumed. 

In the  `weakly' intermittent regime (I2-I6), the density interface is broader ($R\approx 2-5$) and this offset appears correlated with unequal entrainment and mixing (i.e. asymmetry) on either side of the $\bar{\rho}=0$ level. Further observation of the vertical profiles in panels~\emph{g-l} reveals that the density is indeed  better mixed above its $0$ level and that the density interface lies below the velocity interface. This is consistent with the fact that the measured duct volume lies nearer the end sitting in the $\rho=1$ reservoir (i.e. on the `left', as sketched in figure~\ref{fig:setup}\emph{a}, see LPL19, table~2 for the precise locations). Assuming that mixing occurs uniformly across the length of the duct, the bottom layer with initial density $\rho=1$ (coming from the `left') has therefore travelled less, and thus experienced less mixing, than the top layer with initial density $\rho=-1$ (coming from the `right'). This slight, but crucial, non-periodicity along $x$ is an important aspect of SID flows, which appears necessary to obtain continuously-forced exchange flows in the laboratory.

In the more strongly turbulent regime (T2 and especially T3), the vertical density and velocity profiles become similar ($\bar{u}(z) \approx \bar{\rho} (z)$), and closer to $\tanh$/linear. The vertical symmetry of $\TT$ flows and their lower thickness ratio $R \lesssim 2$ result from a more intense and sustained mixing than in $\II$ flows. 

%\AL{Maybe need to read and cite \cite{billant_self-similarity_2001}??}

\subsection{Reynolds averaged balances} \label{sec:reynolds-avg}

We now explain the quasi-steady maintenance of these mean flows $\bar{u},\bar{\rho}$ by analysing the steady Reynolds-averaged $x$-momentum and density equations:
\begin{subeqnarray} \label{RANS}
  \underbrace{\ - \, \overline{\p_x p} \,  }_{\substack{\text{mean pressure} \\ \text{gradient} \, \equiv \, \Pi }}   \underbrace{+ \ Ri_b^s \, \sin \theta \,  \bar{\rho}}_{ \text{body force}} \   \underbrace{+ \ (Re^s)^{-1} (\p_{yy} \bar{u} + \p_{zz} \bar{u})}_{\text{molecular diffusion}}  \ \underbrace{-\ \p_y(\overline{u'v'}) -  \p_z (\overline{u'w'})}_{\text{turbulent diffusion}} \approx 0, \qquad \slabel{RANS-u} \\  
\underbrace{\ - \, \overline{\p_x(u \rho)} \, }_{\substack{\text{mean advective} \\ \text{buoyancy flux} \, \equiv \, \Lambda}} \,  \underbrace{+ \ (Re^s \, Pr)^{-1}  \p_{zz} \bar{\rho}}_{\text{molecular diffusion}} \  \ \underbrace{-\ \p_y(\overline{v'\rho'}) - \p_z (\overline{w'\rho'})}_{\text{turbulent diffusion}}  \approx 0, \qquad \slabel{RANS-rho}
\end{subeqnarray}
where flow fluctuations are defined as $\phi' \equiv \phi - \bar{\phi}$. We used incompressibility $\p_x u +\p_y v + \p_z w = 0$ (imposed at all times) and the (good) approximations that $\bar{v},\bar{w},\p_{yy} \bar{\rho}\approx 0$ and that mean flows are steady (i.e. $\overline{\p_t u} \approx \overline{\p_t \rho} \approx 0$).

The slight but important non-periodicity of the flow in the $x$ direction gives rise to two previously-mentioned key forcing terms: the mean streamwise pressure gradient denoted $\Pi(y,z) \equiv -\overline{\p_x p} $, and the mean streamwise advective buoyancy flux denoted $\Lambda(y,z) \equiv -\overline{\p_x (u \rho)}$ (continuously replacing partially-mixed fluid in the duct by unmixed fluid from the reservoirs).

\begin{figure}
\centering
\includegraphics[width=1.02\textwidth]{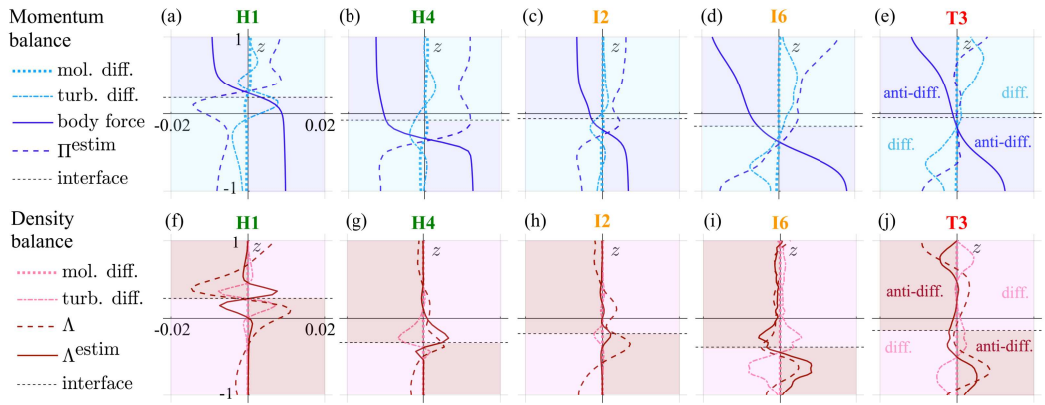}
    \caption{Vertical profiles of the $y$-averaged terms in the Reynolds-averaged \emph{(a-e)} $x$- momentum balance (see \eqref{RANS-u}) and \emph{(f-j)} density balance (see \eqref{RANS-rho}) in five representative data sets H1, H4, I2, I6, T3. Legends and axes limits are identical in all panels.  The horizontal black dashed lines denote the $0$ value of the underlying mean profiles $\langle\bar{u}\rangle_y,\langle \bar{\rho}\rangle_y$ (plotted in figure~\ref{fig:mean_flows}\emph{b,e,g,k,p}). The colouring of quadrants indicate regions of diffusion and anti-diffusion of these profiles (see \emph{e,j} for the legend). `Unexpected' behaviour occurs where line colour does not match quadrant colour.}
    \label{fig:reynolds-avg}
\end{figure}

Figure~\ref{fig:reynolds-avg} shows the vertical structure of each term in \eqref{RANS-u} (top row) and \eqref{RANS-rho} (bottom row) for five representative data sets spanning the $\HH, \II$ and $\TT$ regimes. Derivatives were computed using second-order-accurate finite differences, and we only plot the $y$-average of all terms, neglecting their (weak) spanwise structure. Note that we cannot measure directly the mean pressure gradient $\Pi$ in panels~\emph{a-e}; instead we plot its indirect estimation $\Pi^\textrm{estim}$ assuming a perfect balance of the three remaining terms in \eqref{RANS-u}. Similarly, although we measured the mean advective buoyancy flux as $\Lambda(z) = (2L_x)^{-1} [ \langle u  \rho \rangle_{y,t} (x=0) - \langle u  \rho \rangle_{y,t} (x=2L_x)]$, we also plot for comparison its indirect estimation $\Lambda^\textrm{estim}$ assuming a perfect balance of the two remaining terms in \eqref{RANS-rho}. 

In this two-layer exchange flow, terms in the momentum balance \eqref{RANS-u} that are positive above the $\bar{u}=0$ level (thin black dashed lines in figure~\ref{fig:reynolds-avg}\emph{a-e}) and terms that are negative below this level are both diffusive in the sense that they tend to weaken the flow in each layer and thus decrease $\bar{u}$. These two `diffusive quadrants' are shaded in light blue in the top row (panels~\emph{a-e}) and the terms that are expected to be diffusive (molecular and turbulent diffusion) have a similar light blue line colour. \emph{Vice versa}, terms that have opposite values on either side of the $\bar{u}=0$ level are anti-diffusive in the sense that they tend to strengthen the flow in each layer and thus increase $\bar{u}$. These two `anti-diffusive' quadrants and the terms expected to be anti-diffusive are coloured purple. We extend this diffusive/anti-diffusive distinction to the density balance \eqref{RANS-rho} and bottom row (panels~\emph{f-j}). As a result, unexpected behaviour occurs in regions where line and quadrant colours do not match, which is the focus of the discussion below. 

First, we see that molecular (laminar) diffusion of momentum and density is negligible in all flows (the lines are barely distinguishable from 0), at least in the shear layer region ($|z|\le 1$). By contrast, turbulent diffusion is important in this region, reaching locally absolute values of order $O(0.01)$, which would be responsible for $O(1)$ changes over $O(100)$ advective time units in the absence of counter-acting mechanisms (i.e. over $O(L_t)$, the total time captured in our data sets). Turbulent diffusion behaves diffusively as expected (i.e. these lines are in the quadrant matching their colour), except in the Holmboe regime where these terms are strikingly anti-diffusive in the vicinity of their respective $\bar{u}=0$ and  $\bar{\rho}=0$ interfaces, and diffusive further away from them (panels~\emph{a,b,f,g}). This means that the fluctuations of Holmboe waves effectively sharpen, or `scour' both the velocity and density interface. This sharpening occurs symmetrically on either side of the interfaces in H1 (panels~\emph{a,f}), and asymmetrically (only above the interfaces) in H4 (panels~\emph{b,g}). This is  consistent with the previously-mentioned fact that H1 sustains symmetric (both upward- and downward-pointing) Holmboe waves, while H4 sustains asymetric (upward-pointing only) Holmboe waves.

Second, the gravitational body force is, as expected,  anti-diffusive almost everywhere (i.e. sustaining $\bar{u}$), except in the regions where velocity and density interfaces are offset (panels~\emph{b,c,d}) as discussed in the previous section. However, an unexpected result of panels~\emph{a-e} is that the estimated mean pressure gradient $\Pi^\textrm{estim}$ is  diffusive almost everywhere. In `offset' regions where $\bar{u} \bar{\rho}<0$, this unexpected pressure gradient may provide an  explanation for the sustained offset of interfaces (fluid forced by the pressure gradient to flow against the natural direction suggested by gravitational forcing). However, in `regular' regions where $\bar{u}\bar{\rho}>0$, this unexpected pressure gradient is contrary to our intuition derived from  horizontal ($\theta=0^\circ$) exchange flows where $\Pi$ is necessarily anti-diffusive, as it is the only forcing sustaining the flow. In exchange flows inclined at even small angles (e.g.  $\theta=1^\circ$ in H1) and thus forced by gravity, our results suggest that the particular equilibrium enforced by hydraulic control in the duct causes this pressure gradient to have the opposite effect, i.e. to be diffusive and slow down the flow, at least throughout most of the shear layer. This is  understood from the fact that hydraulic control enforces a $\Delta U\propto \sqrt{g'H}$ velocity scaling (inertial-hydrostatic balance), instead of the  much larger $\sqrt{g'H}\sin\theta Re^s \approx \sqrt{g'H}\theta Re^s$ expected in an infinitely long or periodic tilted duct (gravitational-viscous balance), as explained by the scaling analysis in LL20 \S~2.3. What is not yet understood is the underlying structure of the pressure field, which must be non-trivial near the ends of the duct to match with the far-field hydrostatic pressure into the reservoirs (because, again, hydrostatic pressure alone suggests an anti-diffusive pressure gradient as explained in \S~\ref{sec:hydraulic-nondim}). %Numerical simulations of the whole geometry (including the reservoirs) resolving the pressure field are likely needed to elucidate this interesting result, which may well be a general feature of all inclined two-layer exchange flows.

Third, we see that the measured $\Lambda$ (not closing the density balance) and the estimated $\Lambda^\textrm{estim}$ (closing the density balance) are only significantly different in H1 and I2  (panels~\emph{f,h}). In H1, turbulent anti-diffusion (scouring) near the density interface requires $\Lambda^\textrm{estim}$ to be (unexpectedly) diffusive, but direct measurement of $\Lambda$ suggests otherwise, which is not presently understood. In I2, $\Lambda^\textrm{estim}$ apparently underestimates $\Lambda$, possibly due to limitations in the spatio-temporal resolution of these measurements (see Appendix~\ref{sec:Appendix-data} and quantification of this effect in LPL19, figure~12). This suggests that in some data sets we could use the measured $\Lambda$ as a proxy for turbulent diffusion of density (rather than the other way around). However, doing so would require trust in $\Lambda$ and in the exact balance of \eqref{RANS-u}, and we have seen above that at least one of these could be questionable (see H1).

%Very surprising result that the pressure gradient is in the opposite direction as what we expected!!!! Will need to change some of the text in preceeding sections where we say it is driven by this pressure gradient and everything. Volume averaged budget didn't allow to see that. Actually there is a mistake in LPL19... mean phi k pre doesnt cancel even in forced flows because they will always be a mean pressure gradient. How does it change the results of LPL19 though...??? Maybe best not talk about it here...

%LPL19 looked at volume averaged energy budgets, here we look in more details at the profile of the terms in reynolds averaged momentum and density equation: more demanding! We could say that LPL19 concluded that there was energy/power balance between molecular+turbulent diffusion and body forcing in a volume averaged sense, but here it seems that in the shear layer alone, this is not sufficient to maintain steady state of momentum... we need a mean pressure gradient which wasn't accounted for in LPL19 (we thought it would be smaller than body forcing). How to reconcile those different views????

%Steady state in momentum and buoyancy/density must come from balance of advection/source term and diffusion... steady state balance not present in most DNS and when they are present, this subtle balance has not been investigated in detail \cite{smith_turbulence_2020}.

%%%%%%%%%%%%%%%%%%%%%%%%%%%%%%%%%%%%%%%%% 

\section{Gradient Richardson number and self-organisation} \label{sec:Rig}

\subsection{Definitions}

%
%\begin{equation}\label{def-Rig}
   % Ri_g (\xx,t) \equiv -Ri_b^s \frac{\p_z \rho}{(\p_z u)^2}.
%\end{equation}
%
%this def is useful for large statistics (maybe even turbulent fraction! when $Ri_g<0$) but not in this section
%

The gradient Richardson number $Ri_g(\xx,t) \equiv N^2/S^2$ is the ratio of the square buoyancy frequency $N^2(\xx,t)\equiv -Ri_b^s \p_z \rho$ to the square of the vertical shear frequency $S^2(\xx,t) \equiv (\p_z u)^2$ (in non-dimensional shear layer units, recalling that $\p_z \bar{u}, \p_z \bar{\rho}<0$ throughout the shear layer).  It gives a pointwise measure of the stability of stratified shear flows, since stratification (high $N^2$) tends to stabilise the flow, whereas shear (high $S^2$) tends to destabilise it.

However, in order to work with more tractable (lower-dimensional and smoother) statistics, we consider instead the buoyancy frequency, shear frequency and the gradient Richardson number based on the mean flow:
\begin{equation}\label{def-Rig-bar-bar}
 \bar{\bar{N}}^2(y,z) = -Ri_b^s \,  \p_z\bar{\rho} \qquad \bar{\bar{S}}^2(y,z) \equiv  (\p_z \bar{u})^2, \qquad
\overline{\overline{Ri_g}}(y,z) \equiv \frac{\bar{\bar{N}}^2}{\bar{\bar{S}}^2}\equiv -Ri_b^s \frac{\p_z \bar{\rho}}{(\p_z \bar{u})^2} .
\end{equation}

Note that we use this `double overbar' notation to avoid confusion with the single overbar notation implying the different quantities $\langle N^2\rangle_{x,t}$, $\langle S^2\rangle_{x,t}$, $\langle Ri_g \rangle_{x,t}$, which are noisier and not discussed here.

\begin{figure}
\centering
\includegraphics[width=0.96\textwidth]{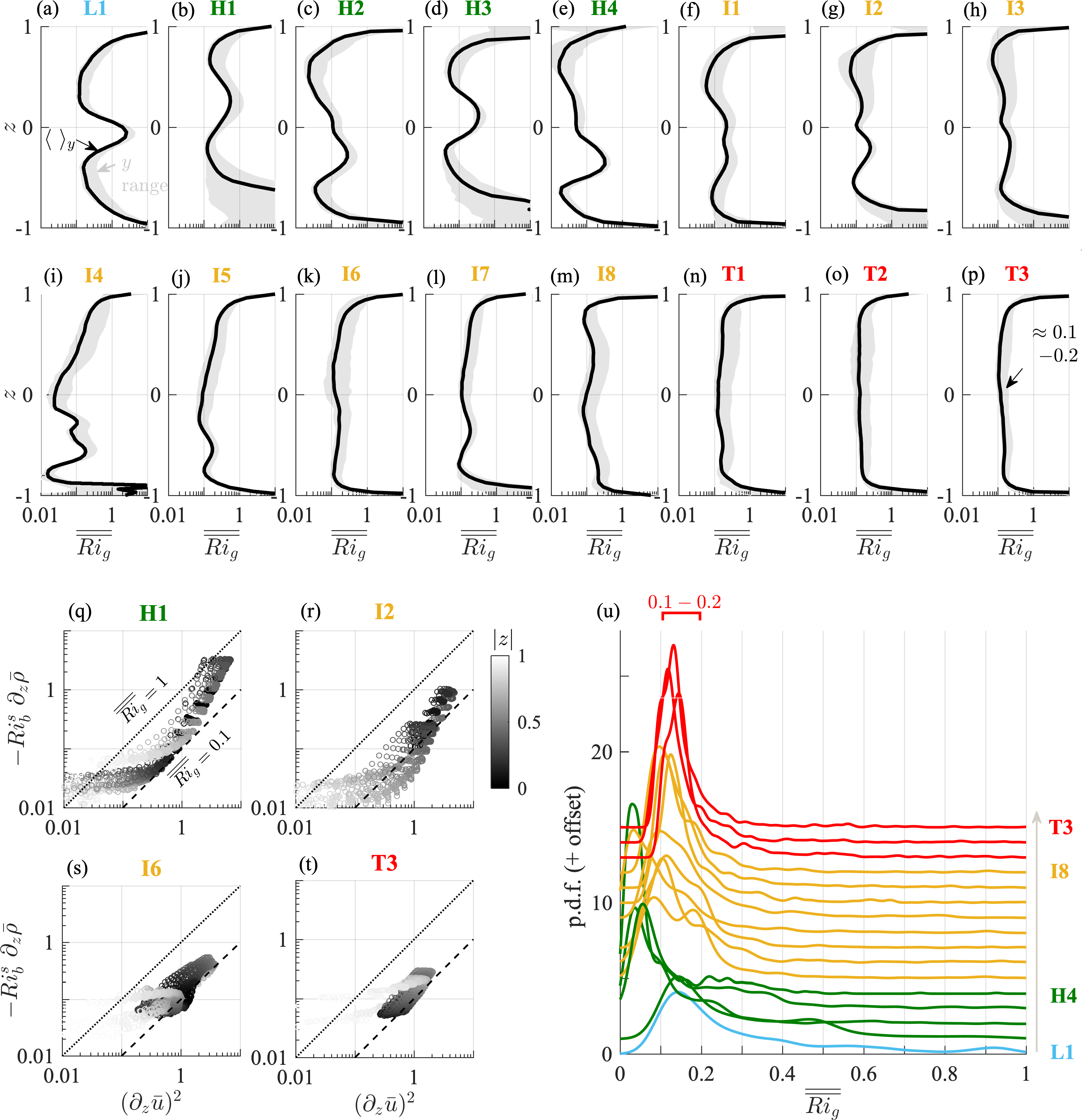}
    \caption{Gradient Richardson number based on the mean flow. \emph{(a-p)} Vertical profiles in all 16 data sets (thick lines denote $\langle \Rigbb \rangle_y$, while grey shadings denote the spread over the entire range $y\in [-L_y,L_y]$). Note the log scale in $\Rigbb$ spanning three decades $0.01-10$. \emph{(q-t)} Correlations  between $\bar{\bar{N}}^2$ and $\bar{\bar{S}}^2$ in H1, I2, I6 and T3. Symbol colours denote the absolute vertical position $|z|$. The dashed and dotted lines correspond to a ratio of $0.1$ and $1$ respectively. \emph{(u)} Probability distribution functions (p.d.f.s)  stacked with successive offsets of $+1$.}
    \label{fig:Rig}
\end{figure}

\subsection{Vertical profiles}

In figure~\ref{fig:Rig}\emph{(a-p)} (first two rows) we plot the vertical structure of this `mean' $\Rigbb$ in all 16 data sets (log-lin scale). We show averages in $y$ across the shear layer (thick black line) together with the total spread across all $y$ locations (grey shading). 

First, we note that the spread in $y$ is generally modest (less than an order of magnitude), especially near the interface ($|z|\lesssim 0.5$). (For a visualisation of the $y$ dependence across the whole duct cross-section in data set T3, see \cite{partridge_versatile_2019} figure~7\emph{(c,i)}). 

Second, focusing on the $y$-averages, we observe that the $\LL$ and $\HH$ profiles (panels~\emph{(a-e}) tend to have two minima of order $0.02-0.1$ on either side of the sharp density interface, and a distinct hump of order $0.2-2$ around the interface. Overall, $\Rigbb$ values tend to monotonically decrease with increasing forcing (i.e. from L1 to T3), except near the edges of the shear layer $z \approx  \pm 1$ where large values are always expected since $\p_z \bar{u}=0$ by definition. 

Third, a clear change in structure occurs in $\II$ profiles, where the single (dromedary) hump of $\LL$ and $\HH$ profiles breaks into a double (camel) hump on either side of the growing interfacial layer of mixed fluid. A final change in structure occurs in the stronger $\II$ and in all $\TT$ profiles, where the double hump flattens and $\Rigbb$ becomes nearly constant at $\approx 0.1-0.2$ across most of the shear layer.

\subsection{Gradient correlations}

In order to understand this last observation that $\Rigbb \rightarrow 0.1-0.2$ in the turbulent shear layer, we investigate in  figure~\ref{fig:Rig}\emph{(q-t)} correlations between the numerator $\bar{\bar{N}}^2$ (vertical axis) and the denominator  $\bar{\bar{S}}^2$ (horizontal axis). We plot, for four representative data sets (H1, I2, I6 and T3), the cloud of all $n_y n_z$ data points (those visible within those axis limits), and denote the absolute $|z|$ position in colour (white representing data near the edges, of lesser interest).

We see in H1 a `comma'-shaped cloud, having a flat low-gradients part corresponding to an asymptote in $\bar{\bar{N}}^2$, and a steep straight high-gradients part corresponding to local values $\Rigbb \approx 0.1-1$ (see the dashed and dotted guide lines). This structure is more or less conserved in I2 (weak $\II$ regime) although lower local values $\Rigbb \ll 0.1$ (below the dashed line) are found at mid-heights (grey colour), in agreement with the profile in panel~\emph{g}. However, very low and very high density gradient  disappear in I6 (strong $\II$ regime) and T3, where the cloud becomes increasingly small and compact around $\bar{\bar{N}}^2\approx 0.04  -0.4$ and $\bar{\bar{S}}^2\approx 0.2-2$, while remaining tangent to the $\Rigbb = 0.1$ scaling (dashed line). %This means that as as the turbulent intensity increases within the shear layer, the mean density and velocity gradients become  tightly linked by a single, near-constant gradient Richardson number.

\subsection{Histograms}

To complement the above observations, we plot in figure~\ref{fig:Rig}(\emph{u}) estimates of the probability density function (p.d.f.) of $\Rigbb$ in all 16 data sets, stacked vertically for visualisation purposes. These p.d.f.s are essentially histograms based on $n_y n_z$ points, normalised such that $\int_0^1 \text{p.d.f.} \, \d \Rigbb = 1$ (note that we ignore the large $\Rigbb>1$ values at the edges of the shear layer). 

This figure shows that the relatively broad and/or multi-peaked p.d.f.s of $\LL$ and $\HH$ flows progressively become narrower and single-peaked in late $\II$ and $\TT$ flows. Intense turbulent flows are thus characterised by mean gradient Richardson numbers  overwhelmingly in the range $0.1-0.2$, with a sharp peak near $0.10-0.15$ in each case.

%Here we don't go into the details of the temporal dynamics of intermittency... rather we treat it as a time-averaged weak form of turbulence 

\subsection{Discussion}

Our $\Rigbb(z)$ data in $\HH$/$\II$ flows bear similarities to the deep-sill ocean overflow data of \cite{van_haren_extremely_2014}, especially to their figure 2\emph{(b)}. They reported long trains of Kelvin-Helmholtz overturning billows in a sustained stratified shear flow with intermittent levels of dissipation  (see  Appendix~\ref{sec:Appendix-litt-rev} for their $Re^s,Ri_b^s$ values). 

Our $\Rigbb(z)$ data in $\TT$ flows are also consistent with the growing body of evidence on the self-organisation of turbulent stratified shear flows subject to `internal mixing', as opposed to `external mixing' imposed by boundary forcing external to the shear layer
\citep{turner_buoyancy_1973}. The evidence suggests that a self-similar equilibrium adjustment of $\bar{u},\bar{\rho}$ occurs such that the gradient Richardson number based on the mean flows is approximately uniform across the shear layer. 

This `equilibrium Richardson number' hypothesis dates back at least to \cite{turner_buoyancy_1973} (see his \S~10.2), who quoted equilibrium values in the literature in the range $\Rigbb(z) \approx Ri_e = 0.06-0.3$.  This hypothesis is also supported by the Monin-Obhukov similarity theory, which assumes a constant buoyancy flux and derives self-similar $\bar{u},\bar{\rho}$ far enough away from any solid boundary (see \cite{turner_buoyancy_1973}, \S~5.1), a regime verified numerically in stratified plane Couette flows
\citep{deusebio_intermittency_2015,zhou_self-similar_2017}. %\pfl{\it I think this is not quite right: MO theory assumes a constant buoyancy flux and derives linear profiles of $u$ and $\rho$ in the strongly stratified limit, where $Ri_e$ applies.} AL: OK, I have modified the text to reflect this

A related `marginal instability' hypothesis was also formulated in \cite{thorpe_marginal_2009} that turbulence maintains itself on the edge of instability flagged by the linear Miles-Howard criterion of $\Rigbb=0.25$, which was supported by the Pacific equatorial undercurrent data and calculations of \cite{smyth_marginal_2013} (see the p.d.f. in their figure 2). A further (related) `self-organised criticality' hypothesis was put forward by \cite{salehipour_self_2018} that strongly-stratified Holmboe wave turbulence is continuously attracted to a critical value of $\Rigbb=0.25$ (see the p.d.f. in their figure 5), making a connection to the scale-invariant energy `avalanches' in the original sand-pile toy model of \cite{bak_self-organized_1988}.

Comparing our $\Rigbb$ data in flows I6-T3 with $(Re^s,Ri_b^s,R,Pr)\approx (10^3, 0.15, 2, 700)$ to the canonical stratified shear layer DNSs of \cite{salehipour_self_2018}  (their figure 13), we find that our `peak' value  $Ri_e \approx 0.10-0.15$ is  lower than their $Ri_e \approx 0.20$ found in `critical Holmboe wave turbulence' with $(Re^s,Ri_b^s,R,Pr)=(6000, 0.16, 10, 8)$, but comparable to their $Ri_e \approx 0.10$ found in `subcritical Kelvin-Helmholtz turbulence' (with much lower $Ri_b^s=0.04, R=1$). However, we note  that their flow is a `run-down' from an initial condition, not forced  as in our experiments.  Comparing to data in  gravity currents forced by a positive $\theta=10^\circ$ slope, our value is compatible with $Ri_e \approx 0.1$ in the  experiments of \cite{krug_turbulent_2015}, figure 8b with $(Re^s,Ri_b^s,Pr) \approx (4000, 0.30, 700)$), but higher than $Ri_e \approx 0.07$ in the DNSs of \cite{reeuwijk_mixing_2019} (see their figure 3a with $(Re^s,Ri_b^s,Pr) \approx (4000, 0.10, 1)$). 

%Although the precise value of this equilibrium Richardson number remains an open question, we believe that our experimental results provide an additional proof that continuously-forced, shear-driven, stratified turbulence tends to self-organise in such an distinctive equilibrium.

% Salehipour 2018:
% All sims at $Re^s=6000$ $Pr=8$

% Critical HWI $Ri_b^s=0.16, R=10$ p.d.f. peaks at $\approx 0.2$

% Sub-critical KHI $Ri_b^s=0.04, R=1$ (weakly-stratified) p.d.f. maximum $\approx 0.1$ 

% Super-critical KHI $Ri_b^s=0.16, R=1$ p.d.f. maximum $\approx 0.35$ 

%Recent paper by Chris H and CPC \cite{howland_testing_2018} where they show Ri=0.25 is a good criterion for suppression of KH billow growth in forced flows. So it does suggest that Ri=0.25 is not just a linear criterion, but something of more profound nonlinear importance in these flows.

%%%%%%%%%%%%%%%%%%%%%%%%%%%%%%%%%%%%%

\section{Turbulent fractions} \label{sec:fraction}

In this section we seek to characterise the distinction between flow regimes in more quantitative and finer ways than done hitherto in ML14, LPL19, and LL20.%, which were largely based on input parameters $\theta,Re^h$. 
We introduce the concept of turbulent fractions, i.e. the ratio of spatial regions that are `turbulent' with respect to two criteria, derived from our simultaneous measurements of the density field and of the three-dimensional, three-component velocity field.
%put some energetics quantitative basis under the definition of the  but here we are looking for some more direct measure, related to enstrophy or density gradient (overturning) as was often used in the literature. This is the concept of turbulent fraction!
We first consider a criterion based on perturbation enstrophy in \S~\ref{sec:fraction-enst}, and then a criterion based on the overturning of the density field in \S~\ref{sec:fraction-overturn}. We then discuss flow visualisations in \S~\ref{sec:fraction-snapshots}, and the dependence on non-dimensional parameters in \S~\ref{sec:fraction-params}. We leave the more detailed statistics on turbulent  energetics to Part 2.% (including the kinetic and scalar energies, their associated fluxes, Fourier spectra, anisotropy, and parameterisations).

% watanabe_turbulent_2016 uses potential enstrophy becaue internal waves contribute total enstrophy (separate turb. from IGW)

\subsection{Perturbation enstrophy fraction} \label{sec:fraction-enst}

We start by defining the perturbation enstrophy as
\begin{equation} \label{def-pert-enstr}
    \omega'^2(\xx,t) \equiv   ||\bnabla \times \uu'||^2,
\end{equation}
where we recall that $\uu' \equiv \uu(\xx,t) - \bar{\uu}(y,z)$ (as defined and used in \eqref{def-avg-1} and \eqref{RANS}). This measure ignores the shear associated with the mean flows (figure~\ref{fig:mean_flows}) in order to capture perturbations away from it, representative of waves or turbulence. Note that the use of our shear-layer-rescaled velocity field (implicit throughout since \S~\ref{sec:shear-layer-rescaling}) ensures that all data sets can be meaningfully compared side-by-side.

\begin{figure}
\centering
\includegraphics[width=1.02\textwidth]{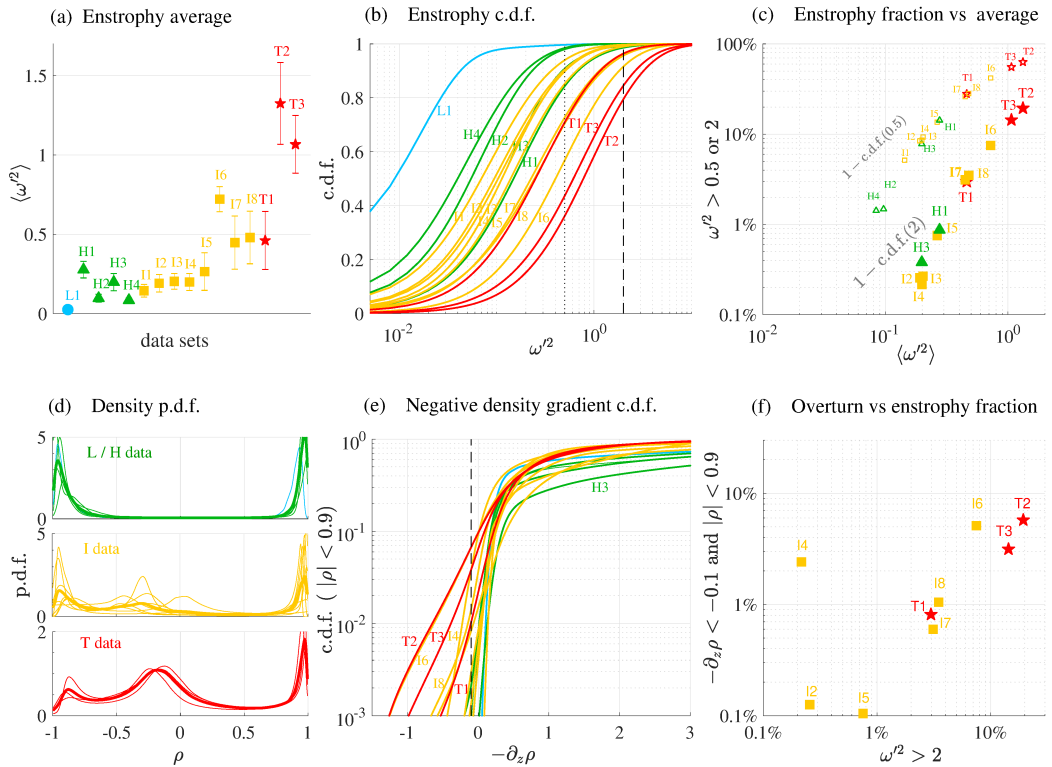}
    \caption{Turbulent fractions based on enstrophy and density overturns in all 16 data sets. \emph{(a)} Time-and volume-average of the perturbation enstrophy $\langle \omega'^2 \rangle$ (error bars show $\pm$ one standard deviation in time of the volume average). \emph{(b)} Cumulative distribution function (c.d.f.) of the perturbation enstrophy (using all $n_x n_y n_z n_t$ points), highlighting the threshold values 0.5 (dotted line) and 2 (dashed line). \emph{(c)} Enstrophy fraction corresponding to threshold values $\omega'^2>0.5$ (small empty symbols) and $\omega'^2>2$ (large full symbols), plotted against the averages of \emph{(a)}. \emph{(d)} Probability distribution function (p.d.f., or normalised histogram) of the density (using all $n_x n_y n_z n_t$ points), separating $\LL$/$\HH$, $\II$, and $\TT$ data for greater clarity. Individual p.d.f.s are shown in thin lines, and the average p.d.f.s for each  sub-panel are shown in thick lines. \emph{(e)} Cumulative distribution function (c.d.f.) of the negative density gradient $\-\p_z \rho$ limited to points where $|\rho|<0.9$, highlighting the `overturn threshold' value $-0.1$ (dashed line). \emph{(f)} Overturn fraction $-\p_z \rho<-0.1$ (with $|\rho|<0.9$), plotted against the enstrophy fraction of \emph{(c)}. Only non-negligible fractions $>0.1\,\%$ are shown. All gradients are computed by second-order finite differences. To remove outliers caused by these gradient computations (for the purpose of this figure only) all of the $\uu'$ and $\rho$ data were smoothed with a spatial filter having an isotropic 3D Gaussian kernel of modest standard deviation of 1 grid point and a tight window of $5\times3\times 5$ grid points in $x,y,z$. }
    \label{fig:enstrophy-overturn}
\end{figure}

First, we plot in figure~\ref{fig:enstrophy-overturn}\emph{(a)} the time- and volume-averaged $\langle  \omega'^2 \rangle$ (as defined in \eqref{def-avg-2}) for all 16 data sets (ordered following the nomenclature of table~\ref{tab:dataset} based on $\theta Re^h$). We also plot the standard deviation in time of this volume average (shown as error bars) to highlight temporal variability. The average $\langle  \omega'^2 \rangle$ increases from $\approx 0$ in L1 to $\approx 0.1-0.3$ in $\HH$ flows, to $\approx 0.2-0.8$ in $\II$ flows, to $\approx 0.5-1.5$ in $\TT$ flows, with some overlap between regimes. This increase is not entirely monotonic; the symmetric Holmboe wave flows H1 and H3 have slightly higher values and temporal variability than the asymmetric Holmboe wave flows H2 and H4, and those values are comparable to the weaker intermittent flows I1-I4, while the stronger intermittent flows I6-I8 are comparable to the weaker turbulent flow T1. Although absolute temporal variability roughly follows a similar pattern, the ratio of standard deviation to mean (sometimes called the coefficient of variation) is fairly constant at $\approx 15-30\,\%$ in most $\HH$, $\II$ and $\TT$ flows, except in I5, I7, I8, T1 where it reaches $\approx 35-45\,\%$. In other words, those four flows could be considered the `most intermittently turbulent', although the remaining $\II$ flows do exhibit turbulent and more quiescent events, and the remaining $\TT$ flows do exhibit temporal variability in the amplitude of their turbulence. Overall, these results confirm the expectation that higher values of $\omega'^2$ and of relative temporal variability, respectively represent higher levels of turbulence and intermittency. Therefore, both provide a quantitative basis generally consistent (but not exactly coincident) with the earlier qualitative flow regime classification.

Second, to go beyond averaged values, we plot in figure~\ref{fig:enstrophy-overturn}\emph{(b)} the cumulative distribution functions (c.d.f.s) of $\omega'^2$, obtained by integration of the p.d.f.s (normalised histograms). All c.d.f.s have a similar sigmoidal shape, with an inflection point at $\approx 0.5$, but their relative position along the $\omega'^2$ axis differs widely, consistent with the pattern of generally increasing $\langle \omega'^2 \rangle$  observed in panel \emph{a}. Moreover, the proximity of some data sets in terms of their averages observed in panel \emph{a} extends to their whole distribution; in particular H3/I2/I3/I4, H1/I5/I7 and I7/I8/T1 have nearly-identical (indistinguishable) c.d.f.s.

Third, we plot in figure~\ref{fig:enstrophy-overturn}\emph{(c)} two sets of enstrophy turbulent fractions, defined as the ratio of data points  above a certain threshold of $\omega'^2>0.5$ (small empty symbols) and  $\omega'^2>2$ (large full symbols) corresponding to $1-\textrm{c.d.f.}(0.5)$ and $1-\textrm{c.d.f.}(2)$, respectively (these thresholds are highlighted by dotted and dashed lines in panel \emph{b}). This enstrophy criterion is loosely based on ideas developed in \cite{holzner_lagrangian_2008,krug_turbulent_2015} for the characterisation of the turbulent/non-turbulent interfaces, and more generally on the fact that turbulence is associated with extreme vorticity fluctuations (long `tail' of the enstrophy p.d.f.s); . These two sets of fractions are plotted against the average values of panel \emph{a}, and reveal an excellent correlation between all three measures. Focusing  on the $\omega'^2>2$ fraction (a threshold value greater than any time- and volume-averages), we find that only the more energetic six data sets I6-T3 have non-negligible fractions $>1\,\%$ representative of significant turbulent events (reaching values of $\approx 20\,\%$ for T2).

%We will use filtered data because of some noise (for a more complete discussion of unfiltered energy and individual velocity/density  spectra see Part 2), but here it s a simplified version.

%Compare this with TKE threshold (probably not the best, also the TKE is only defined latter in the paper. Also compare with density field threshold (maybe the gradient, or N2)...

%Only sensible to use these turbulent fration criteria after the shear layer rescaling (partly why we did it!)

\subsection{Density overturn fraction} \label{sec:fraction-overturn}

Before investigating density overturns, we plot in figure~\ref{fig:enstrophy-overturn}\emph{(d)} the p.d.f.s of the full density field $\rho$, segregating the $\LL$/$\HH$ data (top sub-panel), $\II$ data (middle sub-panel), and $\TT$ data (bottom sub-panel). Individual p.d.f.s (thin lines) and sub-panel averages (thick lines) show a similar trend: $\LL$/$\HH$ flows have a roughly bi-modal distribution $|\rho| \approx 0.9-1$; $\II$ flows develop an extra middle peak flanked by two flat and $\approx 0$ intermediate plateaus; and $\TT$ flows strengthen and broaden the middle peak and increase the value of the intermediate plateaus. This increasingly broad distribution of the density field (from an initially  bimodal $\rho=\pm 1$ distribution in the external reservoirs) owes to the increasing intensity of mixing. Furthermore, the asymmetry of the middle peak, almost systematically between $-0.5<\rho<0$ rather than around $0$, reveals a stronger/more efficient mixing above the density interface ($\rho<0$) than below it. This is consistent with our observations on the mean density profiles in figure~\ref{fig:mean_flows}, which we explained in  \S~\ref{sec:mean_flows} by the non-periodicity of the flow along $x$ and the asymmetrical location of our measuring volume with respect to the duct length. However, note that the c.d.f.s (not shown here) corresponding to these p.d.f.s  are not mathematically equivalent to the mean density profiles $\langle \bar{\rho}\rangle_y(z)$ of figure~\ref{fig:mean_flows}; instead the c.d.f.s of $\rho$ at any given time $t$ would 
yield the instantaneous background density field used to calculate the background potential energy of the flow  \citep{winters_available_1995}, which is beyond the scope of this paper.

Since turbulent mixing is caused by a combination of large-scale stirring and small-scale diffusion, we proceed by investigating in figure~\ref{fig:enstrophy-overturn}\emph{(e)} the c.d.f.s of the vertical gradients of density $-\p_z \rho$ (the negative sign is added for convenience). (Note the use of  lin-log axes in this panel~\emph{e}, as opposed to the log-lin axes in panel~\emph{b}, preventing a direct comparison of the shapes of the c.d.f.s between these two panels). Because of inherent noise in the density field, aggravated by the computation of gradients, we restricted these c.d.f.s to points where $|\rho|<0.9$, i.e. where the density field was at least partially mixed. We find that $\HH$ flows (particularly H3) tend to have sharper stable gradients $-\p_z \rho\gg 1$ (their c.d.f. plateaus to 1 at higher values than most $\II$ and $\TT$ flows). However, $\II$ and $\TT$ flows (particularly I4, I6, I8, T1-T3) tend to have much more unstable gradients $-\p_z \rho< 0$ (left of the dashed line), which signal density overturns. This is consistent with higher levels of turbulent mixing and the earlier qualitative flow regime classification.

Finally, we plot in figure~\ref{fig:enstrophy-overturn}\emph{(f)} the overturn turbulent fraction defined as the ratio of data points having $-\p_z \rho<-0.1$ and $|\rho|<0.9$ (corresponding to $\textrm{c.d.f.}(-0.1)$). These thresholds were chosen to avoid noisy and spurious gradient values caused either by clearly unmixed fluid ($|\rho|> 0.9$) or by very-well mixed fluid ($|\rho|<0.9$ but $\p_z\rho \lesssim 0$). This overturn criterion is loosely based on ideas developed in \cite{portwood_robust_2016} for the identification of dynamically distinct regions in stratified turbulence. The overturn fraction is plotted against the enstrophy fraction of panel~\emph{c}, and the axis limits $>0.1\,\%$ hide the least turbulent flows of lesser interest. Overall, overturn fractions tend to be fairly low ($<6\,\%$), and lower than enstrophy fractions. Moreover, we find a very good correlation between both fractions (most points follow a linear scaling), with the exception of I4, whose overturn fraction is an order of magnitude above that expected (based on its enstrophy fraction, and on the neighbouring flows I2, I5 which we recall have very similar $\theta Re^h$ values). 

%We will seek to understand the relation between turbulent fraction and non-dimensional parameters in \S~\ref{sec:fraction-params}, after discussing flow visualisations in  \S~\ref{sec:fraction-snapshots}.

% Preliminary thoughts... ignore

%Need to show some snapshots of turbulent flows, and turbulent fraction in space. Pretty turbulent flow pictures are important for turbulent paper.

%Talk biefly about difference between I and T flow... basically quantify the fact that intermittent flow are intermittent. Doesn't need to be super precise but gotta say something about it!

%Say that intermittency behaviour (pulsing etc) is for another paper... too much here! Commented section toward end of paper to keep for next paper.

%Plot turbulent fraction with error bar showing the spread in time to emphasise that I is more intermittent than T

%Look at the PDFs/histogram of density profile: the wider it is, the more mixed the flow (related to $\bar{K}_\rho$, maybe in nother section. Well, that's basically what $\bar{K}_\rho$ measures: see energetics section

\subsection{Flow visualisations} \label{sec:fraction-snapshots}

\begin{figure}
\centering
\includegraphics[width=1.03\textwidth]{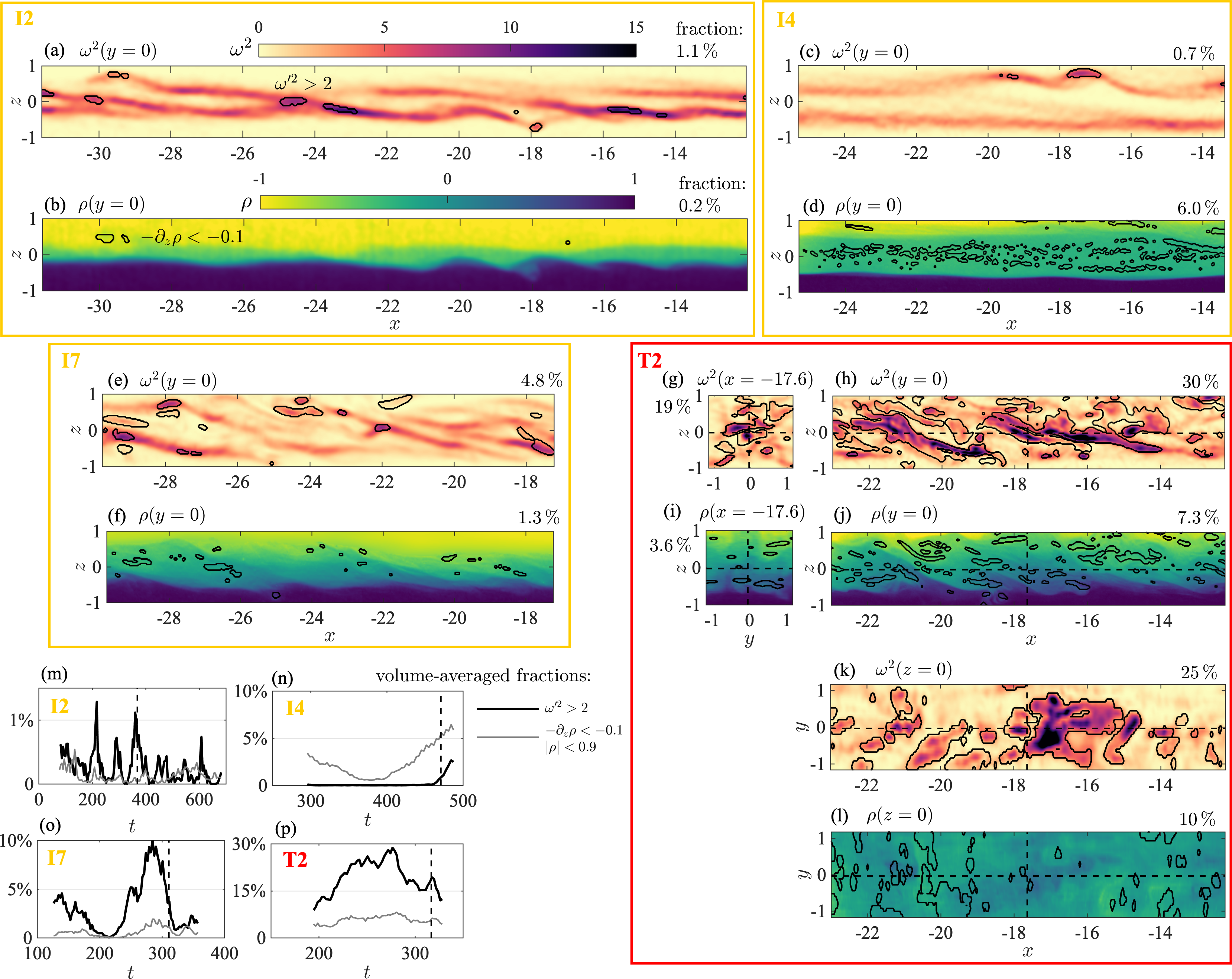}
    \caption{Snapshots and time series of turbulent fractions based on perturbation enstrophy and density overturn for  \emph{(a-b,m)} I2, \emph{(c-d,n)} I4, \emph{(e-f,o)} I7,  and \emph{(g-l,p)} T2. For each data set, we show a single snapshot in time of the \emph{(a,c,e,h)} mid-plane total enstrophy $\omega^2(y=0)$ and \emph{(b,d,f,j)} mid-plane  total density $\rho(y=0)$. For T2 only, we also show the mid-planes \emph{(g,i)} $x=-17.6$ and \emph{(k,l)} $z=0$ (the dashed lines in \emph{(g-l)} denote the location of these plane cuts). Black contours show the regions exceeding the respective turbulent thresholds, and their respective fractions in each plane are given in \%.  For each data set we also show in \emph{(m-p)} the time series of the volume-averaged fractions (the dashed lines denote the time of the snapshots in \emph{(a-l)}). Time does not start at $t=0$ because all data sets were cropped to remove any early-time net flow oscillations as explained in \S~\ref{sec:datasets}. Although the determination of turbulent regions and fractions is based on gradients computed on smoothed $\uu',\rho$ fields (see caption of figure~\ref{fig:fractions_snapshots}), the underlying  $\omega^2,\rho$ snapshots plotted here in colour are not smoothed. Colour bars are identical for all panels.}
    \label{fig:fractions_snapshots}
\end{figure}

To delve deeper into the above observations,  figure~\ref{fig:fractions_snapshots} offers visualisations of these turbulent fractions in four data sets I2, I4, I7 and T2, which are representative of the four main clusters in figure~\ref{fig:enstrophy-overturn}\emph{(f)}. We plot three types of information. First, for each data set (highlighted by the three yellow and one red boxes around panels~\emph{a-l}) we plot a snapshot of the underlying full enstrophy $\omega^2 \equiv ||\bnabla \times \uu||^2$ and the simultaneous density $\rho$ in the mid-plane $y=0$ (for T2 only, we also plot the mid-planes $z=0$ and $x=-17.6$). Second, we identify in black contours the regions exceeding the perturbation enstrophy threshold $\omega'^2>2$ and the overturn threshold $-\p_z\rho <-0.1$ (with $|\rho|<0.9$) as discussed in figure~\ref{fig:enstrophy-overturn}.  The corresponding  turbulent fractions in each plane (relative area in \%)  are displayed in the top right corner of each panel.  Third, we plot in panels~\emph{m-p} the time series of these turbulent fractions averaged over the whole volume (recall that the time-average of these two series was shown in figure~\ref{fig:enstrophy-overturn}\emph{(f)}). The vertical dashed lines in panels~\emph{m-p} denote the time of the respective snapshots in panels~\emph{(a-l}), proving that our choice of snapshots represents typical  (rather than extreme) values. We recall that the mean flows corresponding to these four data sets were shown previously in figure~\ref{fig:mean_flows}\emph{(g,i,l,o)}. We now describe each flow in turn to highlight their salient features.

\subsubsection{I2 flow}

First, we recall that I2 corresponds to the `bottom left quadrant' of figure~\ref{fig:enstrophy-overturn}\emph{(f)} (like I1, I3, I5), representing intermittent flows with the lowest enstrophy and overturn fractions $\sim 0.1\,\%$. 
The enstrophy field (figure~\ref{fig:fractions_snapshots}\emph{(a)}) roughly exhibits two sets of vertically-stacked, quasi-periodic `tilde-shaped' structures (primarily due to $\p_z u$) coinciding with the top and bottom edges of a partially mixed layer in the density field  (figure~\ref{fig:fractions_snapshots}\emph{(b)}). These structures are  spatially only weakly `turbulent' in the sense that the enstrophy only deviates significantly from its long-time mean in a few small regions (in this plane a typical $1.1~\%$), either due to the local weakening or strengthening of the `core' shear or to the shedding of top and bottom `filaments', occasionally coinciding with  limited density overturns (in this plane a typical $0.2~\%$). This weak turbulence is also temporally intermittent (alternating with quiescent periods), as evidenced by the time series in panel~\emph{m}. 

The enstrophy structures (representative of weaker $\II$ flows) can be described as more disorganised and intermittent cousins of the longer-lived tilde-shaped vorticity structures previously described in H4 by \cite{lefauve_structure_2018} (named `confined Holmboe wave') and more generally found in H1-H4. The density field of early $\II$ flows, as compared to $\HH$ flows, also typically features a thicker layer of mixed fluid and interfacial waves of larger amplitude more likely to overturn.

\subsubsection{I4 flow}

Second, we recall that I4 is the only flow in the `top left quadrant' of figure~\ref{fig:enstrophy-overturn}\emph{(f)} having low enstrophy fraction $\sim 0.2\,\%$ 
but medium overturn fraction $\sim 2\,\%$. 
The enstrophy field of I4 (figure~\ref{fig:fractions_snapshots}\emph{(c)}) is generally of lower amplitude than that of I2, and regions exceeding the turbulent threshold remain very limited. The density field of I4 (figure~\ref{fig:fractions_snapshots}\emph{(d)}), exhibits a thicker intermediate mixed layer than that of I2, but with weaker gradients ($|\p_z \rho|\approx 0$) causing widespread (but weak) overturns (in this plane $6.0\,\%$). I4 largely lacks the large-amplitude interfacial waves found in I2 on either edges of the mixed layer, which would normally be associated with perturbation enstrophy (through baroclinic torque), and which are ultimately required to mix the density field by entrainment. 

The time series in panel~\emph{n} gives a clue to explain the apparent paradox of how `so much' mixing (here overturn fraction) could be achieved with relatively `so little' stretching or rotation (here enstrophy fraction). Until $t\approx470$ (the time at which the snapshots are shown), there is indeed very little correlation between both fractions; the overturn fraction undergoes large oscillations while the enstrophy fraction remains close to zero. 

Combining all this evidence on I4, we conclude that the majority of the mixing in I4 likely occurred in vigorously turbulent regions (large enstrophy and overturn fractions) located outside of the measurement volume, and that mixed fluid was subsequently advected into the more quiescent measurement volume (note that the measurement volume spans only $13~\,\%$ of the duct length along $x$). This is consistent with our prior (unpublished) shadowgraphs observations of $\II$ flows along the whole length of the duct, which occasionally showed strong spatial intermittency, i.e. the coexistence and alternation of quiescent and vigorously-turbulent pockets along $x$.

\subsubsection{I7 flow}

Third, I7 represents the intermediate flows in figure~\ref{fig:enstrophy-overturn}\emph{(f)} (like I8, T1) having medium enstrophy and overturn fractions $\sim 1\,\%$. 
The enstrophy field of I7 (figure~\ref{fig:fractions_snapshots}\emph{(e)}) exhibits similar sets of vertically-stacked tilde-shaped structures to that of I2, but these are more disorganised and likely to break off and locally exceed the $\omega'^2>2$ threshold (in this plane $4.8\,\%$). The dynamics of these structures has been described in some qualitative detail in \cite{lefauve_waves_2018} \S~3.3.1, using planar ($y=0$) 2D-2C PIV/LIF measurements at high temporal resolution (see his figure 3.13). Essentially, a turbulent `event' is typically initiated by a small defect in the lower (sharper) density interface, which grows and causes the interface to roll up. The corresponding `single' tilde-shaped vorticity ($\p_z u$) structure (typical of $\HH$ flows) is then stretched by the mean shear, until it eventually splits into two smaller vertically-stacked structures. These structures are in turn stretched and split to create vorticity at finer scales, until the flow is clearly `turbulent'. The corresponding density interfaces undergo successive stretching, ejection of fluid blobs and creation of thin filaments, which promote significant mixing, without large overturns (in this plane only $1.3\,\%$).

The time series in panel~\emph{o} highlights the intermittent character of such events, and confirms that the enstrophy and density snapshots were chosen towards the end of a turbulent event (see the vertical dashed line), after most of the initial stretching and splitting. Furthermore, the good correlation between the enstrophy and overturn fractions is consistent with the local dynamics summarised above (as opposed to the time series of I4 for $t<450$, which required us to invoke advection of mixed fluid from outside the volume).

\subsubsection{T2 flow}

Fourth, T2 corresponds to the `top right quadrant' in figure~\ref{fig:enstrophy-overturn}\emph{(f)} (like I6, T3) having the highest enstrophy and overturn fraction $\sim 1-10\,\%$. 
The enstrophy field of T2 (figure~\ref{fig:fractions_snapshots}\emph{(h)}) has generally much higher values than that of I7 (even locally exceeding the colour bar limits) and thus higher enstrophy fraction ($30\,\%$ in the $y=0$ plane).  The enstrophy field of sustained turbulent flows such as T2 is also much more disorganised than that of I7; tilde-shaped structures are barely detectable among the smaller-scale transient structures. The density field  (panel~\emph{j}) has higher-amplitude and more frequent roll-ups resulting in higher overturn fraction ($7.3\,\%$ in the $y=0$ plane) and in stronger mixing. 
The time series in panel~\emph{p} show that high turbulent fractions ($\gg 1\,\%$) are sustained, despite some unsteadiness, and that both fractions are somewhat correlated.

Finally, $\TT$ flow structures are highly three-dimensional, as evidenced by the cross-sectional $y-z$ cut (panels~\emph{g,i}) and horizontal $x-y$ cut  (panels~\emph{k,l}). A consequence of this three-dimensionality is the high variability of turbulent fractions and the lack of correlation between them in individual planes. For example, the enstrophy fraction is $19\,\%$ in panel~\emph{g} but $30\,\%$ in panel~\emph{h}, while the overturn fraction is $3.6\,\%$ in panel~\emph{i} but $10\,\%$ in panel~\emph{l}. This highlights the importance of three-dimensional, simultaneous data for the study of stratified turbulence in the laboratory.

We refer the reader interested in further three-dimensional visualisations of $\TT$ flows to \cite{partridge_versatile_2019}. Their figures~8-9 show a snapshot of $u,v,w,\omega^2,\rho$ in three perpendicular cuts for flow T3, and include the whole duct cross-section $(y^h,z^h) \in [-1,1]^2$ (i.e. not only the shear layer $(y,z) \in [-1,1]\times [-L_y,L_y]$ as in the present paper).

\subsection{Role of  parameters}
\label{sec:fraction-params}

Now that we have described in more details the turbulence in $\II$-$\TT$ flows, we seek to clarify its relation to non-dimensional parameters. The question is: how could the distribution of turbulent fractions in roughly four clusters in figure~\ref{fig:enstrophy-overturn}\emph{(f)} %(I2, I5; I4; I7, I8, T1; and I6, T2, T3)  
be predicted from the maps of input or output parameters in figure~\ref{fig:output_params}\emph{(a-c)}? 

To mention only a few apparent paradoxes prompting this question: the turbulent fractions of I4-I8 are scattered among all four clusters despite having a nearly equal product of input parameters $\theta Re^h \approx 80-85$ (where $\theta$ is in radians, as in all scaling laws); T1 is much `less turbulent' than T2 and T3 despite having similar input $\theta Re^h \approx 120-130$ and the largest output $Re^s$ of all flows; and I7 is much less turbulent than T2 despite having similar output $Re^s,Ri_b^s,R$. 

The power law regressions of figure~\ref{fig:output_params}\emph{(d-f)} demonstrated the major role of $\theta$ in the (approximate) scaling of the three output parameters $Re^s,Ri_b^s,R$. It is also natural to expect that $\theta$, as key non-dimensional parameter in the governing equations \eqref{eq-motion}, also plays a major role in the turbulent fractions, which is not captured by $Re^s,Ri_b^s,R$ alone.

\begin{figure}
\centering
\includegraphics[width=1.03\textwidth]{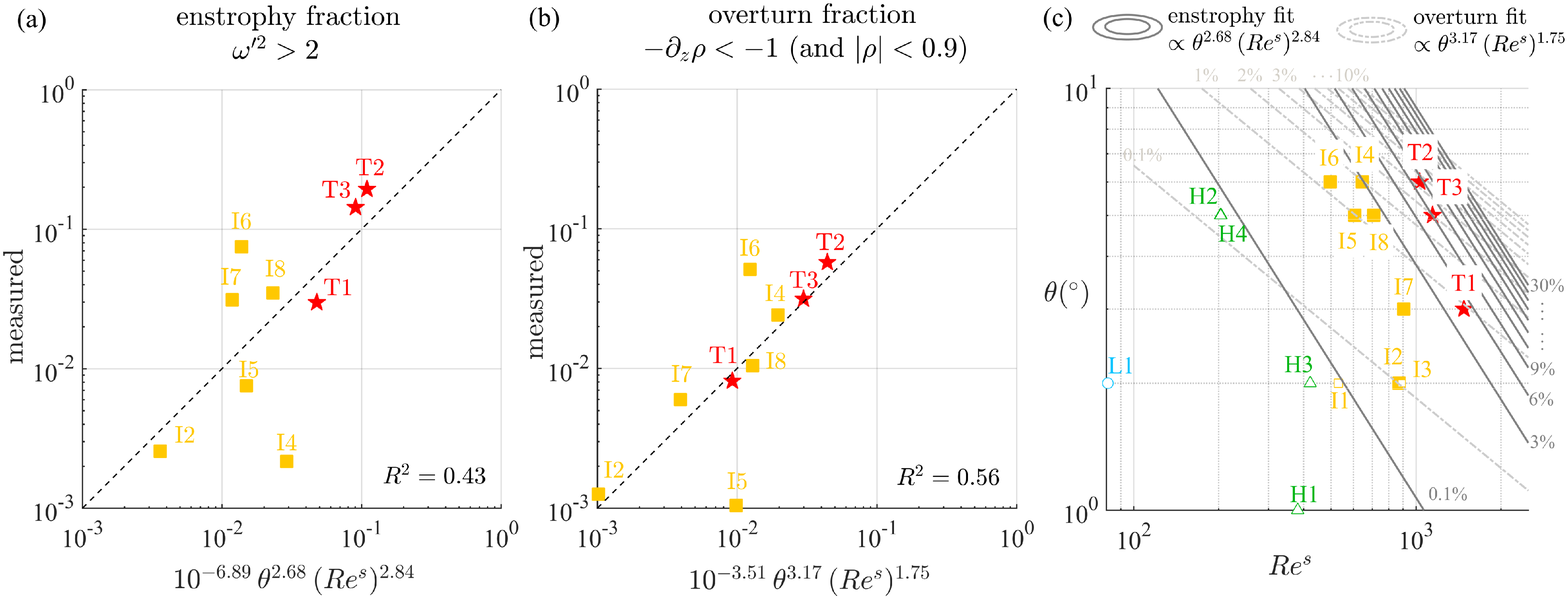}
    \caption{Power law fit of turbulent fractions with $\theta$ and $Re^s$ (least-squares linear regression in log-log space) for the data with $>10^{-3}=0.1\,\%$ fraction of figure~\ref{fig:enstrophy-overturn}\emph{(f)}. \emph{(a)} Enstrophy fraction, and \emph{(b)} overturn fraction (fit \emph{vs} measured value, dashed line denoting equality). \emph{(c)} Map of data sets in the $(\theta,Re^s)$ plane and contours of the enstrophy fraction and overturn fraction fits shown in  \emph{(a,b)}. We plot the $0.1\,\%$ contour, and contours with increments of $3\,\%$ for the enstrophy fraction and of $1\,\%$ for the overturn fraction. Data sets with $<0.1\,\%$ fraction (not used for fitting) are plotted as small open symbols.}
    \label{fig:turb_frac_fits}
\end{figure}

To confirm this, we plot in figure~\ref{fig:turb_frac_fits} the power law regression (best fit) of the turbulent fractions with respect to $\theta$ and $Re^s$, using the nine data sets with turbulent fractions $>0.1\,\%$ plotted in figure~\ref{fig:enstrophy-overturn}\emph{(f)}. A multivariate regression including the additional two parameters $Ri_b^s$ and $R$ was performed, but it provided little additional predictive power, probably because these two parameters are fairly constant across all nine data sets.

First, we see in figure~\ref{fig:turb_frac_fits}\emph{(a)} that the enstrophy fraction follows an approximate scaling $\propto \theta^{2.7} (Re^s)^{2.8}$, and we see in figure~\ref{fig:turb_frac_fits}\emph{(b)} that the overturn fraction follows an approximate scaling $\propto \theta^{3.2} (Re^s)^{1.8}$. This shows that both turbulent fractions increase  steeply (super-linearly) with both $\theta,Re^s$, at least in the `low-fraction' ($\lesssim 20\,\%$) regions investigated here.  This also highlights the fact that the enstrophy fraction scales equally strongly with  $\theta$ and $Re^s$, while the overturn fraction scales more strongly with $\theta$ than $Re^s$. This latter finding is somewhat consistent with the power law regression results of figure~\ref{fig:output_params}\emph{(a,c)} from which we deduce $R\propto \theta^{-1.65}(Re^s)^{0.7}$ i.e. mixing scales more strongly with $\theta$ than with $Re^s$. 
%Remembering from figure~\ref{fig:output_params}\emph{(d)} that $Re^s \propto \theta^0.7 (Re^h)^{1.4}$, we can deduce the corresponding scalings on input parameters as $\propto \theta^{4.8} (Re^h)^{4.0}$ and  $\propto \theta^{4.5} (Re^h)^{2.5}$  respectively. 

Second, we show in figure~\ref{fig:turb_frac_fits}\emph{(c)} the contours resulting from the two turbulent fraction scalings in the $(\theta,Re^s)$  log-log plane (enstrophy fraction fit in solid dark lines, and overturn fraction fit in dashed-dotted light lines), together with the location of data sets in this plane. We see that the fits correctly predict the $0.1\,\%$ fraction `thresholds' since all data sets used for the fitting (full symbols) are indeed located to the right of both $0.1\,\%$ contours (the remaining data sets are shown as small empty symbols, and are located  to the left of at least one $0.1\,\%$ contour). These two sets of superposed contours also illustrate the existence of different clusters (or quadrants) in figure~\ref{fig:enstrophy-overturn}\emph{(f)}: I4, I6 have low $Re^s$ and high $\theta$, and thus achieve relatively high overturn fractions with respect to the enstrophy fraction, and \emph{vice versa} for I7. This can be further quantified by the ratio of overturn-to-enstrophy fraction which follows a scaling $\propto \theta^{0.5} (Re^s)^{-1.1}$, increasing with $\theta$ and decreasing with $Re^s$, and hinting at a general change in  the `type' or `flavour' of stratified turbulence.

Finally, we note that the above empirical scaling relations rely on relatively poor fits (figure~\ref{fig:turb_frac_fits}\emph{(a-b)} have $R^2=0.43$ and $0.56$ respectively). Therefore, these relations have been used to generate qualitative insight rather than  detailed quantitative predictions. They would benefit from being verified by further experimental data sets across a broader range of parameters, and from receiving a theoretical basis.

\section{Conclusions}

%We didn't try to make the data fit in one box and illustrate a particular 'pet' model or idea, but just LOOK AT THE DATA, as it is. Not trying to just show a small interesting shiny rock/one aspect of the flow to fit they pet idea but gives partial view, but just show the whole thing to understand some underlying trend. We did have to make choices given the available spaces but tried to focus on the key things using a non-dimensional  framework and methodology making it easy to compare data sets between each other and with the rest of the literature. Below are the main conclusions that are of relevance to the wider community interested in experimental measurements of continuously forced stratified turbulence and its comparison with DNS and field data:

In this Part 1 we presented some key `basic' properties of continuously-forced, shear-driven, stratified turbulence generated by exchange flow in a square duct inclined at a small angle $\theta$ (SID experiment). We analysed 16 data sets of the simultaneous density field and three-component velocity field in a three-dimensional sub-volume of the duct, spanning a range of non-dimensional parameters and flow regimes. In \S~\ref{sec:methodology} we adopted a convenient shear layer  non-dimensional framework to focus on the core of the flow (discarding near-wall data), and to consistently define the effective `shear layer' Reynolds number $Re^s$, bulk Richardson number $Ri_b^s$, and interface thickness ratio $R$. This allowed for easy comparison of flow profiles and statistics across all data sets, as well as with other results on stratified shear layers to support the three-pronged (observational, numerical, experimental) effort outlined in \S~\ref{sec:context}. Below we summarise the progress made  on the three sets of questions raised in the end of \S~\ref{sec:intro}.

In \S~\ref{sec:mean_prop} we described the non-trivial mapping of SID flows from the space of input parameters $\theta,Re^h,Ri_b^h$ (set by the experimentalist or the numericist) to the space of output parameters $Re^s,Ri_b^s,R$ (set by the internal flow dynamics). We also highlighted that our flows sustain turbulence at much lower $Re^s$ than in unforced stratified shear layer simulations due to the forcing tilt angle $\theta$ and the continuous advection of unmixed fluid from the external reservoirs into the duct.  Next, we investigated vertical profiles of the mean flows and of the Reynolds-averaged equations sustaining them. In particular, we found that Holmboe wave  fluctuations actively sharpen (or `scour') the density interface on which they rely, whereas intermittently turbulent and fully turbulent flows actively broaden it. %\pfl{\it It may be worth commenting on internal and external mixing}  %I have now included a brief discussion of internal vs external mixing earlier in the text (Section 5), and added reference to "internal mixing" to the conclusion and abstract (together with the Rig results,where I think it is more directly relevant
We also discovered the wide-reaching influence of the large-scale streamwise inhomogeneity of the flow (or non-periodicity in the $x$ direction). Some of its effects were readily understood, e.g. the sharpening of the density interface by advection of unmixed fluid, or the asymmetric entrainment and mixing on either side of the interface along the duct. However, some of its effects remain unexplained, e.g. the density and velocity mid-points being substantially offset in some flows, or the role of the mean hydrostatic pressure gradient which  decelerates the flow at $\theta>0$. Numerical simulations resolving the pressure field in the whole geometry (duct and external reservoirs) would help elucidating this question, which might be generic to two-layer, hydraulically-controlled exchange flows inclined at an angle $\theta>0$.

%Interesting bits of figure 4 is that forcing mechanisms are still unclear, some are paradoxical, and the role of non-periodidicity is crucial if we want to understand these flow and compare them to DNS in periodic BCs. First because non-periodicity creates an important refreshing (sharpening of the density), which would have to be mimicked in DNS somehow (e.g.  relaxing the density field to initial value as in \cite{smith_turbulence_2020}), because how else can you sustained stratified turbulence in a periodic volume? 

%Second because non-periodicity creates a hydrostatic pressure gradient which should accelerate the flow, but based on our data (reynolds avg, assuming the budget closes) it seems to slow down the flow in some $\theta>0$ cases. Why is that? Need to investigate pressure BCs in this geometry, perhaps using a LES in the whole geometry including the reservoirs.

%Also we say in figure 4 that Holmboe waves seem to sharpen the density profile (scouring), anti-diffusive effect...

In \S~\ref{sec:Rig} we showed that the vertical profiles of the mean gradient Richardson number $\Rigbb(z)$ smoothly evolved from a single-hump structure due to the strongly-stratified interface of Holmboe flows, to a double-hump structure due to increased mixing in intermittent flows; and finally to a broad plateau $\Rigbb(z)\approx 0.1-0.2$ across most of the shear layer in fully-turbulent flows. As the turbulent intensity increases within the shear layer, we showed that the mean density gradient (buoyancy frequency) and velocity gradients (shear frequency) become tightly linked by a single, near-constant gradient Richardson number, and the probability distribution function of $\Rigbb$ becomes narrowly peaked around $\approx 0.15$. Our data are consistent with prior theories of `equilibrium Richardson number', `marginal stability' or `self-organised criticality', and thus provide further evidence that continuously-forced, shear-driven, stratified turbulence tends to self-organise in such a distinctive `internal mixing' equilibrium. However the precise value of this equilibrium Richardson number differs across the observational, numerical and experimental studies cited, and thus remains an open question.

In \S~\ref{sec:fraction} we quantified the differences between flow regimes by analysing their enstrophy, density and density gradient statistics. We defined two distinct and complementary turbulent fractions as the relative flow volume exceeding a threshold in perturbation enstrophy, or experiencing density overturning. This divided intermittently and fully-turbulent flows into roughly four clusters based on their location in this enstrophy$-$overturn turbulent fraction space, and we investigated these differences using spatial and temporal visualisations of representative flows. This revealed two particular challenges in extracting converged turbulent statistics from our experimental data, acquired over a finite, inhomogeneous sub-volume of the duct and over a finite time period. First, intermittently turbulent flows show cycles with various periods, some exceeding a hundred advective time units (of the order of our recording time). Second, well-mixed turbulent fluid can be suddenly advected into our sub-volume causing  spikes in turbulent fraction, not due to  internal dynamics, but rather to the advection of spatially-intermittent turbulent patches (typically along $x$, but possibly along $y$ and $z$ too since we excluded near-wall data). Despite these challenges, approximate scaling relations between the turbulent fractions and the two key non-dimensional parameters $\theta,Re^s$ suggest that turbulence at high $\theta$ (though here $\theta<10^\circ$) and low $Re^s$  is subject to more overturning and mixing but less extreme enstrophy compared to turbulence at low $\theta$ and high $Re^s$. %(note that $Ri_b^s\approx 0.1-0.2, R\approx 2$ remain fairly constant).

\vspace{0.5cm}

\noindent \textbf{Acknowledgements}

\vspace{0.2cm}
A. L. is supported by an Early Career Fellowship funded by the Leverhulme Trust and the Isaac Newton Trust. We also acknowledge past funding from EPSRC under the Programme Grant EP/K034529/1 `Mathematical Underpinnings of Stratified Turbulence' (MUST) and current funding from the European Research Council (ERC) under the European Union's Horizon 2020 research and innovation Grant No 742480  `Stratified Turbulence And Mixing Processes' (STAMP). Finally, we are grateful for the invaluable experimental support and expertise of Stuart Dalziel, Jamie Partridge and the technicians of the G. K. Batchelor Laboratory.  

\vspace{0.5cm}

\noindent \textbf{Declaration of Interests}

\vspace{0.2cm}

 The authors report no conflict of interest.

\clearpage

\appendix{
%\numberwithin{figure}{section} % so the figures can be called A.1 B.1 etc and not as if they were part of the main document
%\numberwithin{table}{section}

\section{Literature summary table}\label{sec:Appendix-litt-rev}

\vspace{1cm}

\rotatebox{90}{%  % rotate the minipage 90 degrees counter-clockwise
\begin{small}
\begin{minipage}{19cm}
\vspace{-0.5cm}
\setlength\tabcolsep{5pt} 
\renewcommand*{\arraystretch}{1.2}

\bigskip  % whitespace between caption and tabular material
\begin{tabular}{P{4.2cm} P{3.2cm} P{2.1cm} P{1.6cm} P{1.6cm} P{1cm} P{3.8cm}}
\xdef\tempwidth{\the\linewidth}
Studies &  Flow type & Forcing & $Re^s$  & $Ri_b^s$ & $Pr$ & Focus \\ 
\midrule 
 \textbf{Field observations}   \\
 \cite{mcpherson_turbulent_2019} & river plume (Doubtful Sound, New Zealand)& freshwater runoff & $O(10^6)$ & $0.1-1$ & $700$ & turbulent length-scales  \\
\cite{tedford_observation_2009} & estuary (Fraser River) & tide   &  $O(10^6)$ & $O(1)$ & $700$ & shear instabilities\\ 
\cite{geyer_mixing_2010} & estuary  \hspace{1cm}  (Connecticut River) & tide   &  $O(10^6)$ & $\approx 2$ & $700$ & mixing, shear instabilities \\  
\cite{van_haren_extremely_2014} & deep sill overflow (Mid-Atlantic ridge) & AABW &  $O(10^7)$ & $\approx 0.4$ & $7$ & KH billows, mixing \\
\cite{smyth_diurnal_2013} & surface/under current (equatorial Pacific) & wind \& sun & $O(10^8)$ & $\approx 1$ & $7$ & turbulent cycles, marginal instability \vspace{0.1cm} \\ 
 \textbf{Numerical simulations}  \\ 
\cite{smyth_anisotropy_2000} & $\tanh$ & none & $250-1250$ & $0.08-0.16$ & $1-7$ & anisotropy \\ 
\cite{mashayek_time-dependent_2013};  \cite{salehipour_turbulent_2015}; \cite{salehipour_diapycnal_2015}; \cite{salehipour_turbulent_2016,salehipour_self_2018} & $\tanh$  & none & $100-12 \, 000$ & $0.01-0.20$  & $1-16$ & turbulent transition, mixing efficiency, Holmboe turbulence, self-organisation \\
 \cite{watanabe_turbulent_2017} & $\tanh$ & none & $300-500$ & $0.06-0.08$ & $1$ & entrainment, turbulent / non-turbulent interface \\
 \cite{zhou_self-similar_2017,zhou_diapycnal_2017} & stratified \hspace{1cm} plane Couette & boundary & $4250 - 280 \, 000$ & $0.01-1$ & $0.7-70$ & mixing, sharpening, Monin-Obukhov \\ 
\cite{smith_turbulence_2021} & $\tanh$  & relaxation & $4000$ & $0.01-0.35$ & 1 & regimes, mixing, overturning / scouring \\\vspace{0.1cm}
 \textbf{Laboratory experiments}  \\ 
\cite{strang_entrainment_2001} & shear layer over dense quiescent layer & disk pump & $O(10^2)$ & $0.1-0.6$ & $700$ & turbulent entrainment \\ 
\cite{odier_fluid_2009,odier_entrainment_2014,odier_stability_2017}  & wall jet/current over dense quiescent layer & initial jet \hspace{0.9cm} \& wall slope & $250-5000$ & $0.1-0.9$ & $700$ & entrainment, mixing, Thorpe length \\
\cite{lefauve_regime_2019,lefauve_buoyancy_2020} &  exchange through long duct & exchange flow \&  slope  & $200- 30 \,000$ & $0.1-1$ & $700$ & regimes, energetics, interface thickness \vspace{0.2cm} \\
\midrule
\end{tabular}
\captionof{table}{Summary of a few relevant studies on shear-driven stratified turbulence discussed in \S\S~1.2-1.4. The shear Reynolds number $Re^s$ and bulk Richardson number $Ri_b^s$ were estimated and/or converted to match our definitions in \eqref{def-Res-Ribs}.} 
\end{minipage}
\end{small}}

\clearpage

\section{Further properties of the data sets} \label{sec:Appendix-data}

\vspace{0.5cm}

\begin{small}
\begin{center}
\def~{\hphantom{0}}
\setlength{\tabcolsep}{7pt}
  \begin{tabular}{lcccccccccccccc}
      Name &   \multicolumn{4}{c}{Shear layer volume} & \multicolumn{4}{c}{Data points} & \multicolumn{3}{c}{Resolution of data} \\ [1pt]  \cmidrule(l{3pt}r{3pt}){2-5} \cmidrule(l{3pt}r{3pt}){6-9} \cmidrule(l{3pt}r{3pt}){10-12} 
        &    $2L_x$ & $2L_y$ & $2L_z$ & $L_t$ & $n_x$ & $n_y$ & $n_z$ & $n_t$ & $\Delta x,\Delta z$ & $\Delta y$ & $\Delta t$ \\ [5pt]
L1  & 20.3 & 2.72 & 2 & 724 & 403 & 24 & 41 & 251 & 0.050 & 0.11 & 2.89 \\[5pt] % m141
H1 &  19.3 & 3.23 & 2 & 232 & 416 & 34 & 44 & 113 & 0.047 & 0.095 & \textbf{2.05} \\ % m135
H2 &  21.4 & 2.78 & 2 & 595 & 444 & 20 & 41 & 293 & 0.048 & \textbf{0.063} & \textbf{2.03} \\ %m125
H3 & 20.1 & 2.95 & 2 & 321 & 450 & 47 & 45 & 96 & 0.045 & \textbf{0.063} & 3.34 \\ % m129
H4 & 22.0 & 2.76 & 2 & 396 & 442 & 21 & 36 & 198 & 0.050 & 0.13 & \textbf{2.00} \\[5pt] % m133
I1 & 24.6 & 3.55 & 2 & 449 & 430 & 46 & 35 & 71 & 0.057 & \textbf{0.077} & 6.32 \\ % m130
I2 & 19.6 & 2.66 & 2 & 604 & 447 & 26 & 46 & 140 & 0.044 & 0.10 & 4.31 \\ % m131
I3 & 20.8 & 2.94 & 2 & 336 & 445 & 26 & 43 & 67 & 0.047 &  0.11 & 5.01 \\ % m132
I4 & 11.9 & 2.46 & 2 & 191 & 414 & 29 & 66 & 60 & \textbf{0.029} & 0.085 & 3.18 \\ % m144
I5 & 15.9 & 2.47 & 2 & 531 & 403 & 22 & 52 & 263 & 0.039 & 0.11 & \textbf{2.02} \\ % m126
I6 & 11.9 & 2.47 & 2 &  55 & 414 & 28 & 68 & 44 & \textbf{0.029} & 0.088 & \textbf{1.23} \\ % m145
I7 & 12.5 & 2.60 & 2 & 231 & 418 & 29 & 65 & 87 & \textbf{0.030} & 0.090 & 2.66 \\ % m146
I8 & 18.7 & 2.57 & 2 & 275 & 446 & 31 & 49 & 90 & 0.042 & 0.083 & 3.06 \\[5pt] % m138
T1 & 11.4 & 2.52 & 2 & 593 & 402 & 30 & 70 & 151 & \textbf{0.028} & 0.084  & 3.93 \\ % m147
T2 & 10.8 & 2.33 & 2 & 133 & 413 & 30 & 75 & 63 & \textbf{0.026} & \textbf{0.073} & \textbf{2.11} \\ % m143
T3 & 16.0 & 2.27 & 2 & 552 & 449 & 31 & 58 & 149 & 0.036 & \textbf{0.073} & 3.70 \\ % m134
\end{tabular}
  \captionof{table}{Further properties of the 16 volumetric data sets used in this paper, complementing table~\ref{tab:dataset}. The volume size $(2L_x,2L_y,2L_z,L_t)$ (in shear layer units) and the data points $(n_x,n_y,n_z,n_t)$ correspond to the `shear layer' region of interest in this paper (cropped in $y,z,t$ from the original data sets of LPL19) as explained in \S~\ref{sec:shear-layer-rescaling}. The resolution of the data is simply $(\Delta x,\Delta y, \Delta z, \Delta t) \equiv (2L_x/n_x, 2L_y/n_y,2/n_z,L_t/n_t)$. Bold values indicate the best resolutions (smallest values).}
\label{tab:dataset-Appendix}
  \end{center}

\end{small}

}

\bibliographystyle{jfm}

\bibliography{AL_references_2020_12.bib}

\clearpage

\tableofcontents

\end{document}